\RequirePackage{fix-cm}
\documentclass[twocolumn, epjc3, comma, sort&compress, natbib]{svjour3}
\bibpunct{[}{]}{,}{n}{}{,} % to get "[numbered]" references from natbib

\journalname{Eur. Phys. J.}

\usepackage{latexsym}
\usepackage{amsmath}
\usepackage{amssymb}
\usepackage{amsfonts}
\usepackage{CJKutf8}

\usepackage[mathscr,scaled=1.15]{urwchancal}
\DeclareFontFamily{OT1}{pzc}{}
\DeclareFontShape{OT1}{pzc}{m}{it}%
{<-> s * [1.15] pzcmi7t}{}
\DeclareMathAlphabet{\mathpzc}{OT1}{pzc}{m}{it}

\usepackage{color}

\usepackage{supertabular}
\usepackage{placeins}
\usepackage{epsfig}
\usepackage{graphicx}

\definecolor{purple}{rgb}{0.5,0,0.5}
\definecolor{blue}{rgb}{0.0,0,0.9}
\definecolor{prdblue}{rgb}{0.133,0.118,0.498}
\usepackage[colorlinks=true, pdfstartview=FitV, linkcolor=prdblue, citecolor= prdblue, urlcolor=prdblue]{hyperref}

\begin{document}
%\begin{CJK}{UTF8}{song}
\begin{CJK*}{UTF8}{gbsn}

\title{$\,$\\[-6ex]\hspace*{\fill}{\normalsize{\sf\emph{Preprint no}.\
NJU-INP 117/26}}\\[1ex]
Kaon Boer-Mulders function using a contact interaction}

\author{Dan-Dan Cheng（程丹丹)\thanksref{NJU,INP}%
    $\,^{,\href{https://orcid.org/0009-0009-6466-4483}{\textcolor[rgb]{0.00,1.00,0.00}{\sf ID}}}$
    %  ddcheng@smail.nju.edu.cn
\and
    \mbox{Minghui~Ding (丁明慧)\thanksref{NJU,INP}%
    $\,^{,\href{https://orcid.org/0000-0002-3690-1690}{\textcolor[rgb]{0.00,1.00,0.00}{\sf ID}}}$}
\and
\\Daniele Binosi\thanksref{ECT}%
    $\,^{,\href{https://orcid.org/0000-0003-1742-4689}{\textcolor[rgb]{0.00,1.00,0.00}{\sf ID}}}$
\and
    Craig D. Roberts\thanksref{NJU,INP}%
       $^{,\href{https://orcid.org/0000-0002-2937-1361}{\textcolor[rgb]{0.00,1.00,0.00}{\sf ID}},}$
}

\authorrunning{Dan-Dan Cheng \emph{et al}.} % if too long for running head

\institute{School of Physics, \href{https://ror.org/01rxvg760}{Nanjing University}, Nanjing, Jiangsu 210093, China \label{NJU}
           \and
           Institute for Nonperturbative Physics, \href{https://ror.org/01rxvg760}{Nanjing University}, Nanjing, Jiangsu 210093, China \label{INP}
           \and
European Centre for Theoretical Studies in Nuclear Physics
            and Related Areas  (\href{https://ror.org/01gzye136}{ECT*}), \\
            \hspace*{0.5em}Villa Tambosi, Strada delle Tabarelle 286, I-38123 Villazzano (TN), Italy
\label{ECT}
\\[1ex]
Email:
\href{mailto:mhding@nju.edu.cn}{mhding@nju.edu.cn} (MD);
\href{mailto:cdroberts@nju.edu.cn}{cdroberts@nju.edu.cn} (CDR)
            }

\date{2026 March 26}
%\date{2026 March 24}  %  draft 1 complete
%\date{2026 March 12}  %  begun

\maketitle

\end{CJK*}

\begin{abstract}
Using a symmetry preserving treatment of a vector\,$\otimes$\,vector contact interaction (SCI), results are delivered for the four kaon transverse momentum dependent parton distribution functions (TMDs), \emph{viz}.\ helicity-independent (HI) and Boer-Mulders (BM) TMDs for the kaon's $u$, $s$ valence degrees of freedom.  In completing this analysis, we are able to deliver insights into, amongst other things, the role played by emergent hadron mass (EHM) phenomena in producing these TMDs; the EHM modulating effect of the Higgs-boson coupling that produces the strange quark current mass; the impact of gauge link models on whether predictions satisfy the positivity constraint that bounds the BM function relative to the HI TMD; and the size of the BM shift and effects thereupon of off-diagonal terms in the associated scale-evolution kernel.
\end{abstract}

\section{Introduction}
\label{sec1}
In attempting to understand and solve quantum chromodynamics (QCD), it is common to desire an approach that connects, so far as is possible, with notions drawn from quantum mechanics.
In particular, it is widely considered useful to work with hadron wave functions that may be interpreted as probability amplitudes.
In order to achieve that, a light-front formulation of the theory is advantageous \cite{Brodsky:1997de}.
Adopting this perspective, then many parton distribution amplitudes and functions -- quantities with a probability interpretation -- may be obtained directly from a hadron's light-front wave function (LFWF), which is expressed in terms of the light-front fractions, $x$, of the hadron's $4$-momentum carried by a given constituent and their associated $2$-momenta, $\vec{k}_\perp$, in the plane perpendicular to the light-front.
It is worth highlighting that a hadron's LFWF can also be obtained via light-front projection of its Poincar\'e-covariant Bethe-Salpeter wave function \cite{tHooft:1974pnl}; see, \emph{e.g}., Refs.\,\cite{Yao:2025xjx, Xiao:2025cqz} for an application of this approach to pion-like systems.
Against this backdrop, hadron transverse momentum dependent parton distribution functions (TMDs), which are of much contemporary interest \cite{Boussarie:2023izj}, are often said to provide a $3$-dimensional, $(x,\vec{k}_\perp)$ -- light-front, image of the hadron, \emph{i.e}., to deliver a sort of hadron tomographic image.

It is expected that next-generation, high-luminosity, high energy accelerators will deliver data that may be used to reconstruct hadron TMDs \cite{Aguilar:2019teb, Andrieux:2020, Arrington:2021biu, Anderle:2021wcy, AbdulKhalek:2021gbh, Quintans:2022utc, Accardi:2026slw}.
To obtain reliable images, such reconstructions will require both high-level phenomenology and reliable theory.  Much \linebreak
needs to be done in order to reach that point.
This is plain because, today, even the unambiguous inference of a given one-dimensional (only $x$-dependent) distribution amplitude (DA), distribution function (DF), or fragmentation function, is a problem that presents many challenges; see, \emph{e.g}., Refs.\,\cite{Holt:2010vj, Aicher:2010cb, Ball:2016spl, Lin:2017snn, Cui:2021gzg, NNPDF:2021njg, Roberts:2021nhw, Cui:2021mom, Lu:2022cjx, Cheng:2023kmt, Yu:2024ovn, Xu:2024nzp, Xing:2023pms, Xing:2025eip}.

In such circumstances, it is worth asking theory to deliver solid predictions for hadron TMDs that can be used to guide experiment and phenomenology.
In this connection, a picture of contemporary successes and challenges is sketched in Ref.\,\cite[Secs.\,6, 7]{Boussarie:2023izj}.
%At this stage, perhaps, it is too much to expect true QCD predictions of TMD pointwise behaviour.
Some important questions are also highlighted in Ref.\,\cite[Table~1]{Arrington:2021biu}, for instance:
is it possible to obtain data that will enable $3$D imaging of the pion and kaon;
what are the ``universal'' characteristics of such TMDs;
and how do these meson TMDs compare with those of the nucleon -- are there any across-hadron universal features?

We will take steps toward answering these questions by extending the pion study in Ref.\,\cite{Cheng:2024gyv} to the kaon, using the same symmetry preserving treatment of a vector\,$\otimes$\,vector contact interaction (SCI).
Introduced in Ref.\,\cite{GutierrezGuerrero:2010md}, the SCI is not a precision tool; but since that first study, it has been refined so that the modern formulation has many merits, such as:
algebraic simplicity;
simultaneous applicability to many systems and processes;
potential for revealing insights that connect and explain numerous phenomena;
and service as a valuable tool for checking the viability of algorithms used in calculations that depend upon high performance computing.
Moreover, contemporary SCI applications are typically parameter-free, with numerous benchmarking predictions being available for a wide range of phenomena involving mesons
\cite{GutierrezGuerrero:2010md, Roberts:2011wy, Chen:2012txa, Serna:2017nlr, Zhang:2020ecj, Xing:2022sor, Cheng:2024gyv, Sultan:2024hep, Xing:2025eip, Gutierrez-Guerrero:2019uwa, Chen:2024emt}
%% Zhang:2020ecj
and baryons
\cite{Gutierrez-Guerrero:2019uwa, Chen:2024emt, Wilson:2011aa, Segovia:2013rca, Xu:2015kta, Yin:2019bxe, Raya:2021pyr, Cheng:2022jxe, Yu:2025fer, Bai:2026nqo}.

Within the past decade or so, numerous studies of pion TMDs have been completed; see, \emph{e.g}., Refs.\,\cite{Lu:2012hh, Pasquini:2014ppa, Wang:2017onm, Ahmady:2019yvo, dePaula:2020qna, Zhu:2023lst, Kou:2023ady}.
Pion properties are dominated by dynamical effects that are closely linked with the three pillars of emergent hadron mass (EHM)
\cite{Horn:2016rip, Roberts:2020udq, Roberts:2021nhw, Ding:2022ows, Ferreira:2023fva, Raya:2024ejx, Achenbach:2025kfx, Binosi:2026tre}; in particular, with the dynamical generation of a running quark mass, $M(k^2)$, which is large at infrared momenta, $M(k^2\simeq 0)\approx m_p/3$, where $m_p$ is the proton mass, but practically negligible at ultraviolet momenta; see, \emph{e.g}.,  Ref.\,\cite[Fig.\,2.5]{Roberts:2021nhw}.
Apart from generating the light-quark current mass, $m_\ell$, which is the seed for a nonzero physical pion mass \cite[Sec.\,5]{Roberts:2020udq}, Higgs boson couplings into QCD have little impact on pion properties.  (We assume isospin symmetry throughout.)

The kaon is different and therefore interesting.
This case involves the $s$ quark/antiquark, for which the Higgs coupling generates current masses, $m_s$, that are $\approx 27$-times larger than the mean light-quark mass that is relevant for pion observables \cite{ParticleDataGroup:2024cfk}.
The natural question is: what impact does this very large current mass imbalance within kaons have on their observable properties?
This is addressed, \emph{e.g}., in Refs.\,\cite{Cui:2020dlm, Roberts:2021nhw}, which explain that, insofar as parton DAs or DFs are concerned, modern continuum Schwinger function methods (CSMs) predict that EHM strongly screens the disparity in current masses.
Thus, regarding valence quark DAs and DFs, instead of impacts on the order of $27/1$, roughly $5/4$ effects should be perceived.
At a basic level, this is because whilst, in the ultraviolet, the running $s$-quark and $u$-quark mass functions do exhibit a ratio $\approx 27/1$, this value drops to $5/4$ in the infrared \cite[Fig.\,2.5]{Roberts:2021nhw}.
Plainly, empirical comparisons between kaon and pion properties provide valuable tests of the EHM paradigm \cite{Horn:2016rip, Aguilar:2019teb, Andrieux:2020, Arrington:2021biu, Anderle:2021wcy, Quintans:2022utc}.

Today, however, there are only a few theoretical studies of kaon TMDs, for instance:
(\emph{i}) helicity independent kaon TMDs are computed in Ref.\,\cite{Zhang:2024adr}, using a framework similar to that which we will employ;
(\emph{ii}) in Ref.\,\cite{Kaur:2020vkq}, Gaussian \emph{Ans\"atze} for LFWFs are used to compute pion and koan TMDs, including the helicity-dependent Boer-Mulders (BM) function \cite{Boer:1997nt};
and (\emph{iii}) basis light-front quantisation is used in Ref.\,\cite{Lu:2026mkb} to estimate the helicity-independent kaon TMD.
Each of these studies has limitations, \emph{e.g}.:
in omitting the BM function, (\emph{i}) and (\emph{iii}) are incomplete;
and, owing to their constituent-quark-like formulation, neither (\emph{ii}) nor (\emph{iii}) can provide any objective link between Higgs couplings into QCD and kaon properties.

Our discussion is organised as follows.
Section~\ref{sec2} sketches some formal material, including definitions and symmetry constraints that are used in the subsequent discussion.
The kaon helicity-independent TMD is described in Sect.\,\ref{sec3}.  It includes the relevant SCI algebraic formulae, numerical results, and comparisons with the analogous pion TMD.  (Indeed, comparisons with pion results are included in every section.)
Section~\ref{sec4} is devoted to the kaon BM function.  Of particular interest is a discussion of the gauge link, whose presence is essential to obtaining a nonzero BM function.
Following a different path, the TMDs are rederived in Sect.\,\ref{LFWFcalc}, namely, the SCI kaon LFWF is calculated and used to deliver TMDs via overlap representations.
Some features of TMD evolution are discussed in Sect.\,\ref{TMDEvolve}.
Of special interest is an exploration of the impact of off-diagonal terms in the evolution equation for the leading $k_\perp^2$ moment of the BM functions.  This section also contains an analysis of the so-called BM shift, which is the mean transverse $y$-direction momentum of $x$-direction polarised valence degrees-of-freedom (dof) in the unpolarisable pseudoscalar meson.
Section~\ref{epilogue} provides a summary and a perspective.

\section{Pseudoscalar Mesons TMDs}
\label{sec2}
%\subsection{Context}
Considering a ``$q$'' valence quark in the kaon, one works with a Dirac matrix valued quark + quark correlation function, $\Phi^{q/K}(x,\vec{k}_\perp)$, detailed below, which has the following two-function decomposition:
\begin{align}
\Phi^{q/K}(x,\vec{k}_{\perp})  & =
\tfrac{1}{2} \left[ f_{1K}^q(x,k_\perp^2) i \gamma\cdot n \right.  \nonumber \\
& \left. \qquad + h_{1K}^{q\perp}(x,k_\perp^2)  \tfrac{1}{4f_K}  \sigma_{\mu\nu} k_{\perp \mu} n_\nu
\right]\,, \label{DF_decomposition}
\end{align}
where
$ f_{1K}^q$ is the helicity-independent kaon TMD and $h_{1K}^{q\perp}$ is the kaon BM function.
The BM term in Eq.\,\eqref{DF_decomposition} is scale-normalised via $f_K$, \emph{i.e}., the kaon leptonic decay constant.
This choice is made because a pseudoscalar meson leptonic decay constant, $f_{\mathsf 5}$, is an order parameter for chiral symmetry breaking \cite{Holl:2004fr}.
Consequently, $f_{\mathsf 5} \neq 0$ in realistic studies of pseudoscalar mesons, which necessarily express dynamical chiral symmetry breaking, itself a corollary of EHM.
Scale normalisation using the pseudoscalar meson mass, $m_{\mathsf 5}$, is common; however, that is problematic in the chiral limit.
In Eq.\,\eqref{DF_decomposition},
$n$ is a lightlike four-vector, $n^2=0$;
$\bar n$ is its conjugate, $\bar n^2=0, n\cdot \bar n= -1$;
and $k_{\perp \mu} = O_{\mu\nu}^{\perp} k_\nu$,
$O_{\mu\nu}^{\perp} = \delta_{\mu\nu}+n_\mu \bar n_\nu + \bar n_\mu n_\nu$.
Of course, ``$q$'' valence-quark in pion TMDs are readily obtained by taking the limit $m_s\to m_\ell$ in any self-consistent calculation \cite{Cheng:2024gyv}.

We would like to highlight that, unless otherwise noted, the results presented herein are to be interpreted as expressing meson structural properties at the hadron scale, $\zeta_{\cal H}<m_p$.
At $\zeta_{\cal H}$, all properties of a given hadron are carried by its quasiparticle valence dof \cite{Yin:2023dbw}.
The existence of such a scale is ensured by the theory of effective charges in QCD \cite{Grunberg:1980ja, Grunberg:1982fw}, \cite[Sec.\,4.3]{Deur:2023dzc}.
A perspective on TMD evolution to higher scales is presented elsewhere \cite{Boussarie:2023izj}.
Given a TMD, both the $x$ and $k_\perp^2$ profiles evolve with scale according to equations that contain elements which are essentially nonperturbative and poorly known at present.
Thus, today, there is practitioner-choice dependence in evolution outcomes.
This is especially true for the BM function.
Herein, the TMD rapidity scale dependence is not shown explicitly.  In practical applications, it is often chosen to be the same as $\zeta$ \cite{Bacchetta:2017gcc}, \emph{viz}.\ $\zeta_{\cal H}$ herein.

The TMD $f_{1K}^q(x,k_\perp^2)$ is a $1+2$-dimensional number density: it expresses helicity-independent $x$-$k_\perp^2$ correlations in the kaon valence quark LFWF as measured by a vector (photon) probe.  The associated term in $\Phi^{q/K}(x,\vec{k}_{\perp})$ is even under action of the time-reversal operator, $T$; hence, nonzero, even in simple formulations of the problem.

Baryon number conservation entails
\begin{subequations}
\label{BNC}
\begin{align}
1 & = \int_0^1 \! dx \int \! d^2\vec{k}_\perp f_{1K^+}^{\bar s}(x,k_\perp^2)\,, \\
& = \int_0^1\!  dx\int \! d^2\vec{k}_\perp f_{1K^+}^u(x,k_\perp^2) \,,
\end{align}
\end{subequations}
with analogous identities for the other kaons and the pions.
\emph{N.B}.\ The left-hand side is $\zeta$-independent.
In addition, at $\zeta_{\cal H}$:
\begin{subequations}
\begin{align}
1
& = \int_0^1 \! dx \, x \int \! d^2\vec{k}_\perp\,  [ f_{1K^+}^{\bar s}(x,k_\perp^2)+ f_{1K^+}^{u}(x,k_\perp^2)] \\
& =: \int_0^1 \! dx \, x [u^{K^+}(x)+\bar s^{K^+}(x)] \\
& = \langle x u \rangle_{\zeta_{\cal H}}^{K^+} + \langle x \bar s \rangle_{\zeta_{\cal H}}^{K^+}  \,,
\end{align}
\end{subequations}
where $u(x)^{K^+}\!$, $\bar s(x)^{K^+}$ are hadron-scale valence dof DFs in the $K^+$.
There are analogous identities for the other kaons and the pions; and this set of constraints states that valence dof carry all hadron momentum at $\zeta_{\cal H}$.

\begin{figure}[t]
\centerline{%
\includegraphics[clip, width=0.40\textwidth]{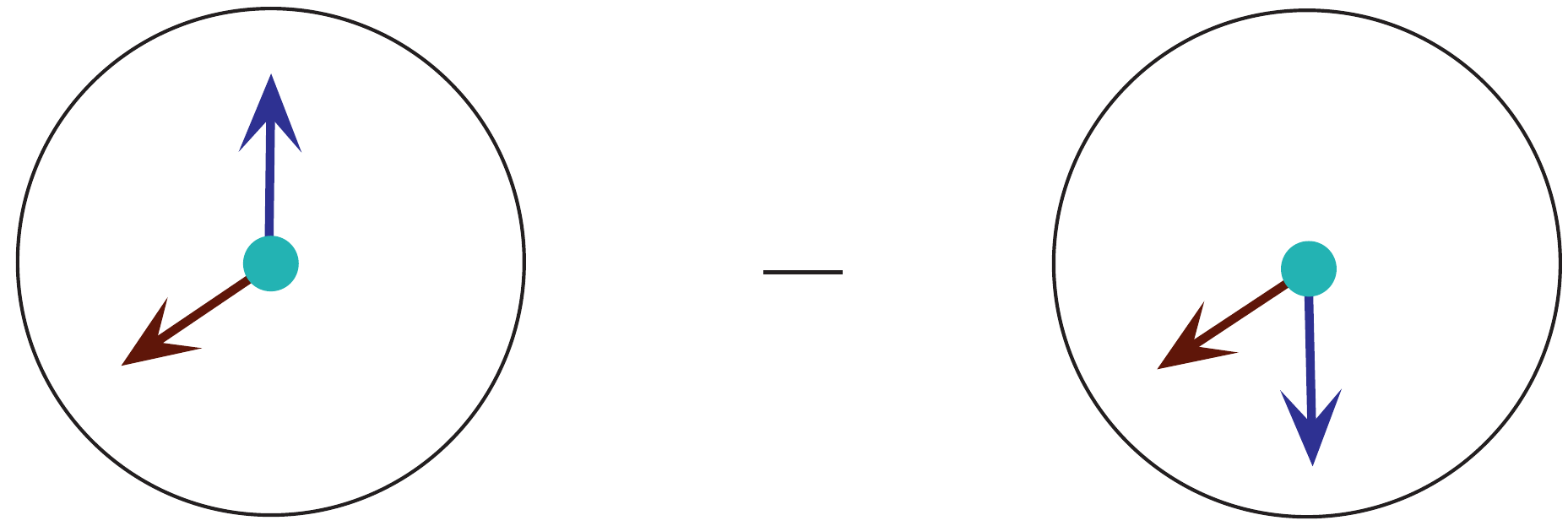}}
\caption{\label{FBM}
Number density interpretation of the Boer-Mulders function.
Legend.  Vertical blue arrows -- transverse polarisation of the quark, $\vec{S}$;
oblique brown vectors -- quark transverse momentum vector, $\vec{k}_\perp$. }
\end{figure}

\begin{figure*}[t]
\centerline{%
\includegraphics[clip, width=0.775\textwidth]{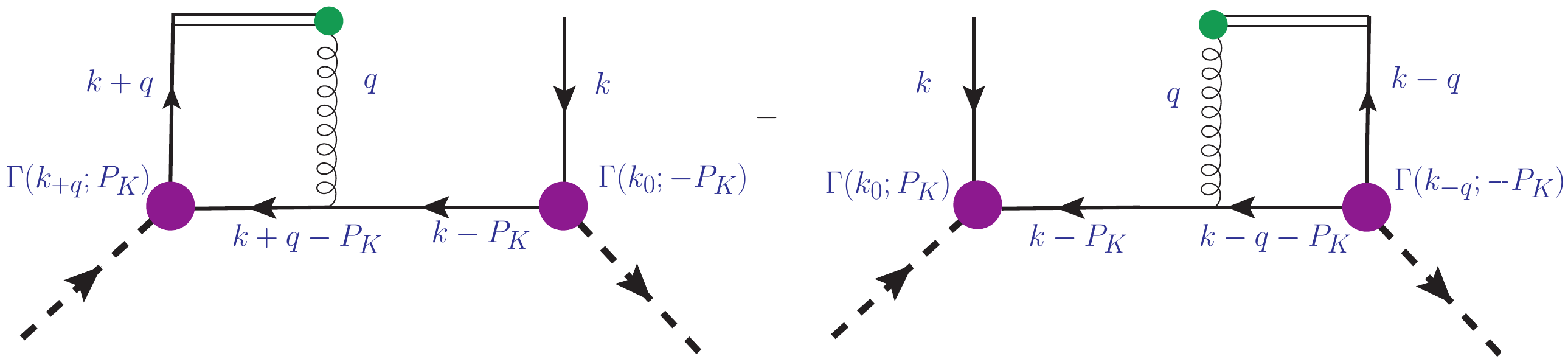}}
\caption{\label{ImageBM}
In order to obtain a nonzero Boer-Mulders function, one must require, at least, that the (valence) dof involved in the forward scattering event subsequently/initially interacts with the spectator via (multiple) gluon exchanges.
Legend.
Double line -- leading eikonal approximation to quark propagation under the influence of the gauge link, $1/n\cdot q$;
small green circle -- eikonal quark-gluon vertex, $[-g n_\mu]$, where $g$ is the strong coupling;
large purple  circle -- kaon (pseudoscalar meson) Bethe-Salpeter amplitude, $\Gamma_K(k_{\pm q,0};P)$, with $k_{\pm q}=k\pm q-P_K/2$, $k_0=k-P_K/2$;
thin solid line -- valence quark propagators, $S(k)$;
spring-like line -- gluon propagator, $D_{\mu\nu}(q)$;
and dashed lines -- incoming/outgoing mesons.
The relative negative sign between the two diagrams expresses the sign-change between initial- and final-state eikonal-quark interactions.
%
%$k_{1}=k+q-\frac{P_{\pi}}{2}$, $k_{2}=k-\frac{P_{\pi}}{2}$; $k_{1}^{\prime}=k-q-\frac{P_{\pi}}{2}$,$k_{2}^{\prime}=k-\frac{P_{\pi}}{2}$ are the relative momenta in pion Bethe-Salpeter amplitude.
}
\end{figure*}

On the other hand, $h_{1K}^{q\perp}(x,k_\perp^2)$ is $T$-odd; so, the origin of a nonzero result is worth recapitulating.
The kaon's valence dof are $J=1/2$ fermions.
Thus, measured with respect to the kaon $3$-momentum, it is possible that the number density distribution of valence quark transverse spins is sensitive to the quark's transverse momentum and that this dependence can be revealed by a vector probe; see Fig.\,\ref{FBM}.
This connection with in-hadron spin correlations makes the BM function a prominent focus of modern phenomenology and theory; and recognition of the pivotal role played by pseudoscalar mesons in elucidating consequences of EHM and its modulation by Higgs boson couplings into QCD means that developing an understanding of kaon and pion BM functions has become especially important.

%\subsection{Helicity independent}
Moving to details, the Dirac matrix valued quark + quark correlation function discussed above has the following general form:
\begin{align}
\label{D-17}
\Phi_{ij}^{q/K}(x,\vec{k}_\perp) & =
\int \frac{d^4\xi}{(2\pi)^3} \delta(n\cdot \xi)
e^{ik\cdot\xi}
\langle K(P)|\bar{q}_j(0) \nonumber  \\
& \quad \times {\cal L}^{\bar n}(0,\infty) {\cal L}^{\bar n}(\infty,\xi)
q_i(\xi)| K(P)\rangle\,,
\end{align}
%% \delta(n\cdot \xi) has dimensions of mass
%% n = {0, 0, 1, I}/Sqrt[2]
%% nb = {0, 0, -1, I}/Sqrt[2]
%% a+ = n.a ... a- = bar n . a
where $i,j$ are spinor indices and $P$ is the kaon total momentum.
%Here we express ourselves with a Minkowski metric, for ease of comparison with existing literature, \emph{e.g}., Ref.\,\cite{Lu:2016pdp}.
%%Thus:
%%$n$ is a lightlike four-vector, $n^2=0$;
%%$\bar n$ is its conjugate, $\bar n^2=0$, $n\cdot \bar  n = 1$;
%%$x=n\cdot k/n\cdot P$, with $k$ the momentum of the active quark and $P$ is the total momentum of the pion;
%%and $(O_{\mu\nu}^\perp k^\nu) = (0,\vec{k}_\perp,0)$,
%%$O_{\mu\nu}^\perp = g_{\mu\nu} - n_\mu \bar n_\nu - \bar n_\mu n_\nu$.
%
The correlator in Eq.\,\eqref{D-17} involves gauge links, $ {\cal L}^{\bar n}$, that ensure gauge invariance: the paths are \cite{Lu:2016pdp} a product of one term running along the negative light-cone between zero and infinity, $\tilde{\cal L}^{\bar  n}$, and another connecting these points along a transverse path, ${\cal L}^{\bar n {\rm T}}$.
In light-cone gauge, for which the $n\cdot A$ component of each gauge field is set to zero, then $\tilde {\cal L}^{\bar  n} \equiv 0$, whereas ${\cal L}^{\bar n {\rm T}}\equiv 0$ in Feynman gauge.
Connecting Eqs.\,\eqref{DF_decomposition}, \eqref{D-17}, one has
\begin{subequations}
\label{TMDprojections}
\begin{align}
f_{1K}^q(x,k_\perp^2) & = {\rm tr} \tfrac{1}{2} i\gamma\cdot\bar n \Phi^{q/K}(x,\vec{k}_\perp) \,,
\label{TMDprojectionA}\\
\frac{k_\perp^2}{4 f_K} h_{1K}^{q\perp}(x,k_\perp^2) &
= {\rm tr}\tfrac{1}{2}  \sigma_{\mu\nu} k_{\perp \mu} \bar n_\nu \Phi^{q/K}(x,\vec{k}_\perp)
\label{TMDprojectionB}\,.
\end{align}
\end{subequations}
%%  16 f^2 ...8 f -> 2 f ... 1
Hereafter, we write ${\mathpzc M}_K = 4 f_K$; similarly for the pion.  Empirically, ${\mathpzc M}_K \approx m_p/2$ \cite{ParticleDataGroup:2024cfk}.

If one neglects the gauge links in Eq.\,\eqref{D-17}, then, as signalled above, the helicity-independent number densities, $f_{1K}^q(x,k_\perp^2)$, are nonzero; however, the BM functions vanish, $h_{1K}^{q\perp}(x,k_\perp^2)\equiv 0$.  (Similarly for $\pi$.)
This highlights that interactions between the spectator of the initial scattering event and the dof struck by the probe are crucial to obtaining $h_{1K}^{q\perp}(x,k_\perp^2) \neq 0$.
Herein, following, \emph{e.g}., Refs.\,\cite{Lu:2012hh, Cheng:2024gyv}, we realise such interactions via the processes sketched in Fig.\,\ref{ImageBM}, \emph{i.e}., by introducing an eikonal representation for the struck quark interaction with the gauge link \cite{Collins:1989gx}.
Notably, in the calculation of $f_{1K}$, as with $f_{1\pi}$, one finds that the two contributions in Fig.\,\ref{ImageBM} cancel; so, this gauge link model yields no contribution to the unpolarised TMD.

Referring to the material above, the number density distribution of valence dof whose polarisation is transverse to the kaon's three-momentum direction vector, $\hat P$, is conventionally defined as follows \cite{Bacchetta:2004jz}:
\begin{align}
f_{q^\uparrow/K}(x,\vec{k}_\perp) & = \tfrac{1}{2} f_{1K}^q(x,k_\perp^2) \nonumber \\
& \qquad - h_{1K}^{q\perp}(x,k_\perp^2)\tfrac{1}{2 {\mathpzc M}_K} \hat P \times \vec{k}_\perp \cdot \vec{S}_q\,.
\end{align}
(For a kaon, it is a fair approximation to consider $P \parallel n$.)  Then the Fig.\,\ref{FBM} number density asymmetry corresponds to the following difference:
\begin{align}
f_{q^\uparrow/K}(x,\vec{k}_\perp) &  - f_{q^\downarrow/K}(x,\vec{k}_\perp) \nonumber \\
& =  h_{1K}^{q\perp}(x,k_\perp^2) \tfrac{|\vec{k}_\perp|}{{\mathpzc M}_K} \sin( \phi_S - \phi_{k_\perp})\,,
\end{align}
where the azimuthal angles are measured between the indicated transverse vector and $\hat{P}$.
Exploiting the requirements imposed by positivity of the defining matrix elements, one readily arrives at the following pointwise positivity bound \cite{Bacchetta:1999kz}:
\begin{equation}
\label{EqPositive}
 |k_\perp h_{1K}^{q\perp}(x,k_\perp^2)/{\mathpzc M}_K| \leq f^q_{1K}(x,k_\perp^2)\,.
\end{equation}

\section{Helicity-Independent Kaon TMD}
\label{sec3}
The helicity-independent kaon TMD can be calculated from the diagrams in Fig.\,\ref{ImageBM} by omitting the gauge link, which also means changing the relative sign from ``$-$'' to ``$+$''.
Focusing on the $u$-in-$K$, this yields:
\begin{align}
 f_{1K}^{u}(x, & k_{\perp}^{2}) =
N_{c}{\rm tr}_{\rm D}\int\frac{d k_{3}dk_{4}}{(2 \pi)^{4}}\delta_{n}^{x}(k) \Gamma_K (- P_K)
\nonumber \\
& \quad \times S_{u}(k)in\cdot\gamma S_{u}(k)\Gamma_K(P_K) S_{s}(k-P_K),
\label{24/9/11.1}
\end{align}
where $N_c=3$, the trace is over spinor indices;
$\delta_{n}^{x}(k)=\delta(n\cdot k-xn\cdot P_{K})$;
$\Gamma_K$ is the kaon Bethe-Salpeter amplitude;
and $S_{s,u}$ are dressed $s, u$ quark propagators, respectively.
Using the Ward identity \cite{Ward:1950xp} and exploiting the fact that the integral of a total derivative is zero, it is readily established that, at $\zeta_{\cal H}$:
\begin{equation}
\label{TMDSymmetry}
f_{1K^+}^{\bar s}(x,  k_{\perp}^{2}) = f_{1K^+}^{u}(1-x, k_{\perp}^{2}) \,,
\end{equation}
with similar identities for the other kaons.

We give mathematical meaning to the diagrams by specifying the quark + quark interaction,
computing every element that appears,
then combining them to produce a numerical result.
For the interaction, we use the SCI, a brief recapitulation of which is provided in \ref{AppendixSCI}.
In this way, one obtains the following result for the kaon helicity-independent TMD:
\begin{align}
f_{1K}^{u}(x, k_{\perp}^{2}) & =
\frac{N_c}{2\pi^3} \left[E_{K}^2\mathcal{N}_{EE} -\frac{(M_{s}+M_{u})E_{K}F_{K}}{M_{s}M_{u}}\mathcal{N}_{EF}\nonumber\right.\\
& \qquad \left.+\frac{F_K^2}{4}\frac{(M_{s}+M_{u})^2}{M_{s}^2M_{u}^2}\mathcal{N}_{FF}
\right]\,,
\label{uinKTMD}
\end{align}
where,
$m_K$ is the kaon mass,
$M_{u,s}$ are dressed-quark masses -- see Eq.\,\eqref{genS},
$E_{K}, F_{K}$ express the SCI kaon Bethe-Salpeter amplitude -- see Eqs.\,\eqref{PSBSAA}, \eqref{kaonBSA},
and
\begin{subequations}
\begin{align}
 \mathcal{N}_{EE} & = \frac{\bar{C}_{2}(\varsigma)}{\varsigma} + \frac{3\bar{C}_{3}(\varsigma) x \check{x}\left[m_K^2 - \check M_{su}^2\right]}{\varsigma^2}\,,\\
 \mathcal{N}_{EF} & = \frac{\bar{C}_{2}(\varsigma)[xM_{u} + \check{x}M_{s}]}{\varsigma}\nonumber\\
&\quad +\frac{3\bar{C}_{3}(\varsigma) x \check{x}(M_{s}+M_{u})\left[m_K^2 - \check M_{su}^2\right]}{\varsigma^2}\,,\\
\mathcal{N}_{FF} & = \frac{\bar{C}_{2}(\varsigma)(\check{x}-x)(M_s^2-M_u^2)}{\varsigma}\nonumber\\
& \quad +\frac{3\bar{C}_{3}(\varsigma) x \check{x}(M_s+M_u)^2\left[m_K^2 - \check M_{su}^2\right]}{\varsigma^2}\,,
\end{align}
\label{HIkaonTMDFX}
\end{subequations}
\hspace*{-0.2\parindent}where
$\check{x} = (1-x)$,
$\check M_{su} = M_s - M_u$,
$\varsigma = k_{\perp}^2 + x M_{s}^{2} + \check x M_{u}^{2} - x \check x  m_K^2$,
for later use,
$\varsigma_{0} = \varsigma|_{k_{\perp}^2=0} = xM_{s}^{2} + \check x M_{u}^{2} - x \check x m_K^2$,
and the special functions $\bar{C}_{n}$ are defined in Eq.\,\eqref{eq:Cn}.
Of course, when one sets $M_s=M_u$ and replaces the kaon Bethe-Salpeter amplitude by the pion amplitude, one recovers the pion result, $f_{1\pi}(x, k_{\perp}^{2})$; see Ref.\,\cite[Eq.\,(8)]{Cheng:2024gyv}.

\begin{figure}[t]
\vspace*{0.4em}

\leftline{\hspace*{0.5em}{\large{\textsf{A}}}}
\vspace*{-2ex}
\includegraphics[width=0.41\textwidth]{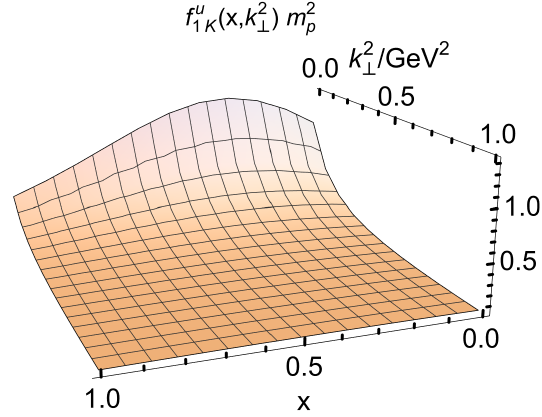}
\vspace*{0.1ex}
\leftline{\hspace*{0.5em}{\large{\textsf{B}}}}
\vspace*{-2ex}
\includegraphics[width=0.41\textwidth]{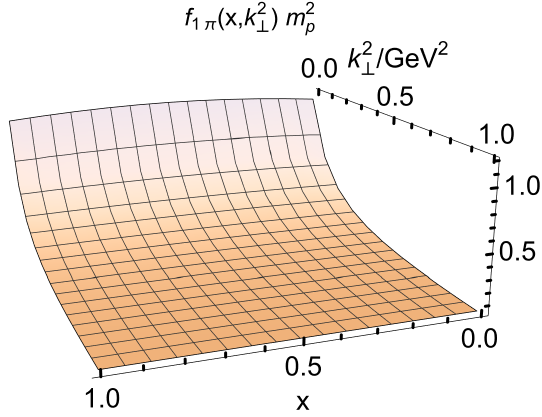}
\vspace*{3.5ex}
\caption{\label{Plotf1}
{\sf Panel A}. Hadron scale SCI result for the helicity-independent $u$-in-$K$ TMD drawn from Eq.\,\eqref{uinKTMD}.  The $s$-in-$K$ TMD is obtained via Eq.\,\eqref{TMDSymmetry}.
{\sf Panel B}. Analogous $u$-in-$\pi$ TMD.}
\end{figure}

In our (standard) formulation of the SCI, described in \ref{AppendixSCI}, Eq.\,\eqref{uinKTMD} yields the kaon TMD drawn in Fig.\,\ref{Plotf1}\,A.
Owing to Eq.\,\eqref{TMDSymmetry}, only the $u$-in$K$ result is drawn.
Unlike the analogous pion TMD, Fig.\,\ref{Plotf1}\,B, the $\zeta_{\cal H}$ $u$-in-$K$ TMD is asymmetric around $x=1/2$: for $k_\perp^2=0$, its maximum value lies at $x \approx 0.3$, \emph{i.e}., it is shifted by roughly $(4/5)^2$; see the Introduction.
(The peak position is weakly sensitive to $k_\perp^2$, something which indicates that a factorised approximation to the kaon LFWF is reasonable \cite{Raya:2021zrz, Yao:2025xjx}.)
%%  This is an indication that a factorised Ansatz is a fair approximation for the LFWF.
The peak relocation owes to Higgs boson couplings into QCD \cite{Cui:2020dlm}.
(For the same reason, $f_\pi/f_K = 0.84$.)
Further, as another consequence of these Higgs couplings, the $u$-in-$K$ $k_\perp^2$-profile shows some dependence on $k_\perp^2$: in this case, $m_K^2/({\mathpzc M}_K^2) \approx 1$, so the kaon mass plays a role.
On the other hand, for the pion, $m_\pi^2/({\mathpzc M}_\pi^2) \ll 1$, hence the $k_\perp^2$-profile is almost $x$-indepen\-dent.

It is also worth noting that, using the SCI, both the kaon and pion TMDs are nonzero on $x\simeq 0,1$ at any finite $k_\perp^2$.
This is an artefact of the momentum-indepen\-dent quark + quark interaction: with an interaction that becomes weaker with increasing momentum transfer, the hadron scale TMD vanishes at these endpoints \cite{Lu:2021sgg}.

Working with the kaon helicity-independent TMD, Eq.\,\eqref{TMDSymmetry}, and using Eq.\,\eqref{IntCbar}, one readily arrives at an expression for the hadron-scale in-kaon valence $u$ quark DF:
{\allowdisplaybreaks
\begin{subequations}
\label{TkaonTMD}
\begin{align}
u^{K}&(x)=\frac{N_{c}}{4\pi^2}\left[E_{K}^2\tilde{\mathcal{N}}_{EE} -\frac{(M_{s}+M_{u})E_{K}F_{K}}{M_{s}M_{u}}\tilde{\mathcal{N}}_{EF}\nonumber\right.\\
&\left. \qquad \qquad +\frac{F_{K}^2}{4}\frac{(M_{s}+M_{u})^2}{M_{s}^2M_{u}^2}\tilde{\mathcal{N}}_{FF}\right]\,,
\label{HIkaonTMD4} \\
\tilde{\mathcal{N}}_{EE} & = \bar{C}_{1}(\varsigma_0)
 + \frac{2\bar{C}_{2}(\varsigma_0)x \check x \left[m_K^2 - M_{su}^2\right]}{\varsigma_0}\,,\\
\tilde{\mathcal{N}}_{EF} & = \bar{C}_{1}(\varsigma_0)  [xM_{u} + \check x M_{s}]\nonumber\\
& \quad +\frac{2\bar{C}_{2}(\varsigma_0) x \check x (M_{s}+M_{u})\left[m_K^2 - M_{su}^2\right]}{\varsigma_0}\,,\\
\tilde{\mathcal{N}}_{FF} & =\bar{C}_{1}(\varsigma_0)  (\check x-x)(M_s^2-M_u^2)\nonumber\\
& \quad + \frac{2\bar{C}_{2}(\varsigma_0) x \check x (M_s+M_u)^2\left[m_K^2 - M_{su}^2\right]}{\varsigma_0}.
\label{HIkaonTMD5}
\end{align}
\end{subequations}
Naturally, pion results are recovered for $s\to d$.

Using the information in \ref{AppendixSCI}, these expressions yield:
\begin{equation}
 \langle x u \rangle_{\zeta_{\cal H}}^{K^+} + \langle x \bar s \rangle_{\zeta_{\cal H}}^{K^+}
 = 0.47 + 0.53 = 1\,.
\end{equation}
Again, the ratio $\langle x \bar s \rangle_{\zeta_{\cal H}}/\langle x u \rangle_{\zeta_{\cal H}} = 1.13$ is typical, being the same in more sophisticated studies \cite{Cui:2020tdf}, and reflects Higgs boson modulation of EHM.
For the pion, the analogous ratio is unity.%
\footnote{We suspect an error was made in computing the helicity-independent kaon DF reported in Ref.\,\cite{Zhang:2024adr} because it yields $\langle x u \rangle_{\zeta_{\cal H}}^{K^+}=0.56> \langle x \bar s \rangle_{\zeta_{\cal H}}^{K^+}=0.44$.  This is physically unreasonable: the lighter valence dof cannot carry a greater light-front fraction of the hadron's momentum than the heavier dof.}
}

\section{Kaon Boer-Mulders Function}
\label{sec4}
\subsection{Formal remarks}
Using Eq.\,\eqref{TMDprojectionB}, the kaon BM function is readily obtained from the mathematical expression that corresponds to Fig.\,\ref{ImageBM}.  In proceeding, it is worth noting that one can first use time-reversal invariance, expressed in the following operations,
\begin{equation}
\label{TRops}
\begin{array}{ll}
{\mathpzc T}^\dagger \gamma\cdot n^{\rm T} {\mathpzc T} & = \gamma\cdot n \,, \\
{\mathpzc T}^\dagger \sigma_{ \alpha +}^{\rm T} {\mathpzc T} & = -\sigma_{ \alpha +} \,,\\
{\mathpzc T}^\dagger S(k)^{\rm T} {\mathpzc T} & = S(k)\,, \\
{\mathpzc T}^\dagger \Gamma_K(k;P)^{\rm T} {\mathpzc T} & = \Gamma_K(k;-P)\,,
\end{array}
\end{equation}
where $(\cdot)^{\rm T}$ denotes matrix transpose; $\sigma_{\alpha +} = \sigma_{\alpha\beta} n_\beta$; and
${\mathpzc T}= \gamma_5 C$, with $C=\gamma_2\gamma_4$ being the charge-conjuga\-tion matrix,
to show that, in general, irrespective of the quark + quark interaction, the second diagram in Fig.\,\ref{ImageBM} maps into the first.  Consequently, the SCI yields:
\begin{align}
h_{1K}^{u\perp}&(x, k_{\perp}^{2}) \frac{\vec{k}_{\perp\alpha}}{4 f_K} =
N_{c}{\rm tr}_{\rm D}\int \frac{d^{4}q}{(2 \pi)^{4}}\frac{d k_{3}dk_{4}}{(2 \pi)^{4}}\delta_{n}^{x}(k)
\nonumber \\
& \times S_{s}(k-P_K ) \Gamma_{K} (- P_K)S_{u}(k)\sigma_{ \alpha +}\frac { [-gn_{\mu }] }{n\cdot q}S_{u}(k +q)\nonumber\\
&\times \Gamma_{K}(P_K)S_{s}(k+q- P_K)[-g\gamma_{v}]D_{\mu \nu}(q)\,,
\label{BMkaon1}
\end{align}
where $1/[n\cdot q]$ is the propagator of the eikonalised quark line and $[-gn_{\mu }]$ is the associated coupling to the gluon \cite{Collins:1989gx}, with $g$ being the strong coupling parameter;
and $D_{\mu \nu}$ is the gluon propagator that mediates the target + spectator interaction.
As with the helicity-independent TMD, at $\zeta_{\cal H}$:
\begin{equation}
\label{BMSymmetry}
 h_{1K^+}^{\bar s\perp}(x,  k_{\perp}^{2}) = h_{1K^+}^{u\perp}(1-x, k_{\perp}^{2}) \,.
\end{equation}
Similar identities hold for the other kaons.

It is worth recalling that, in our Euclidean metric formulation, the struck-quark on-shell condition for semi-inclusive deep inelastic scattering (SIDIS) is expressed via $1/[n\cdot q] \to - \pi \delta(n\cdot q)$, whereas $1/[n\cdot q] \to + \pi \delta(n\cdot q)$ for Drell-Yan (DY).  Using Eq.\,\eqref{BMkaon1}, one then has, alike with the pion:
\begin{equation}
h_{1 K}^{\perp}(x, k_{\perp}^{2})_{\rm SIDIS} =
- h_{1 K}^{\perp}(x, k_{\perp}^{2})_{\rm DY}.
\end{equation}

\subsection{Gauge link}
\label{subsecGL}
Working with the SCI, using Eqs.\,\eqref{KDinteraction}\,--\,\eqref{KCI} to write
\begin{equation}
g^2 D_{\mu\nu}(q) = \frac{4\pi \alpha_{\rm IR}}{m_G^2} \delta_{\mu\nu}\,,
\label{SCIglue}
\end{equation}
then the SIDIS form of Eq.\,\eqref{BMkaon1} evaluates to the following expression:
\begin{subequations}
\begin{align}
\tfrac{1}{{\mathpzc M}_K} h_{1K}^{u\perp~\text{MI}}&(x, k_{\perp}^{2})  =
- \frac{\alpha_{\rm IR}}{m_{G}^2} \frac{ N_{c}}{16\pi^3}\frac{\bar{C}_{2}(\varsigma)}{\varsigma} \left(E_{K}^2\mathcal{M}_{EE}\nonumber\right.\\
&\left. \qquad + E_{K}F_{K}\mathcal{M}_{EF}+F_{K}^2\mathcal{M}_{FF}\right)\,, \\
\mathcal{M}_{EE}& = 4 \bar{C}_{1}(\varsigma_{0})\left[xM_{s}+\check x M_{u}\right],\\
\mathcal{M}_{EF} & =
-\frac{2(M_{s}+M_{u})}{M_{s}M_{u}} \bigg[ \bar{C}_{1}(\varsigma_{0}) (\check x(M_{u}^2+xm_{K}^2) \nonumber\\
& \qquad +2M_{s}M_{u} +xM_{s}^2) - C_{0}(\varsigma_{0})\bigg]\,,\\
\mathcal{M}_{FF} & = \frac{(M_{s}+M_{u})^3}{M_{s}^2M_{u}^2} \bigg[
 - C_{0}(\varsigma_{0})
 \nonumber \\
& \qquad + \bar{C}_{1}(\varsigma_{0}) (M_{s}M_{u} + x \check x m_{K}^2)\bigg]\,,
\end{align}
\label{KaonMIBMFFX}
\end{subequations}
\hspace*{-0.2\parindent}Evidently:
the BM function is only nonzero because of the gauge-link mediated interaction between the spectator and the eikonalised quark -- this is signalled by the $\alpha_{\rm IR}/m_G^2$ factor.
Regarding the chiral limit, the BM function is nonzero so long as the associated pseudo\-scalar meson is a Nambu-Goldstone boson,
and the magnitude of the effect reflects the scale of EHM; see the factors of ${\mathpzc M}_K$, $M_{u,s}$.

Referring to Ref.\,\cite{Cheng:2024gyv}, one expects that the BM function in Eq.\,\eqref{KaonMIBMFFX} is incompatible with the positivity bound, Eq.\,\eqref{EqPositive}.  Direct calculation confirms this; see Sect.\,\ref{SecPositivity} below, namely, with a momentum-independent interaction used to calculate the gauge link contribution, positivity is violated.
This is typical of treatments that provide greater support to the gauge link contribution than they do to the quark loop in the helicity-independent case -- compare, \emph{e.g}., Refs.\,\cite{Lu:2004hu, Lu:2005rq}; and also the studies in Refs.\,\cite{Pasquini:2014ppa, Wang:2017onm, Ahmady:2019yvo, Kou:2023ady}, which violate positivity for similar reasons.

To proceed, therefore, we following Ref.\,\cite{Cheng:2024gyv} in modifying the definition of the gauge link.
Namely, we consider a more realistic representation of the gluon line:
\begin{equation}
\label{momentumglue}
g^2 D_{\mu\nu}(q) = \delta_{\mu\nu} \frac{4\pi \alpha_{\mathpzc L}}{q^2 + m_G^2}\,.
\end{equation}
Taking $\alpha_{\mathpzc L} = \alpha_{\rm IR}$, then Eq.\,\eqref{momentumglue} reproduces the pure-SCI form at infrared momenta.  On the other hand, for any value of $\alpha_{\mathpzc L}$,  Eq.\,\eqref{momentumglue} provides damping in the ultraviolet:
{\allowdisplaybreaks
\begin{subequations}
\label{KaonMDBMFFX}
\begin{align}
 \tfrac{1}{{\mathpzc M}_K} h_{1K}^{u\perp}& (x, \boldsymbol{k}_{\perp}^{2})
 = - \frac{\alpha_{\mathcal{L}}N_{c}}{16\pi^3} \frac{\bar{C}_{2}(\varsigma)}{\varsigma}\int_0^1d\upsilon\left(\tilde{\mathcal{M}}_{EE}E_{K}^2\nonumber\right.\\
& \qquad \qquad \left.+\tilde{\mathcal{M}}_{EF}E_{K}F_{K}+\tilde{\mathcal{M}}_{FF}F_{K}^2\right)\,,\\
%\label{KaonMDBMF}
%
\tilde{\mathcal{M}}_{EE}& = - 8(\upsilon-1)\left[xM_{s}+\check xM_{u}\right] \frac{\bar{C}_{2}(\tilde{\varsigma})}{\tilde{\varsigma}}\,,\\
\tilde{\mathcal{M}}_{EF} & = \frac{2(M_{s}+M_{u})}{M_{s}M_u}
\bigg[ \bar{C}_{1}(\tilde{\varsigma}) \nonumber\\
& \quad +  2 (\upsilon-1)
(x \check x m_{K}^2 + xM_{s}^2+2M_{s}M_{u} \nonumber \\
& \quad + \check x M_{u}^2 + \upsilon k_{\perp}^2)
\frac{\bar{C}_{2}(\tilde{\varsigma})}{\tilde{\varsigma}}\bigg] \,, \\
\tilde{\mathcal{M}}_{FF} & = - \frac{(M_{s}+M_{u})^3}{M_{s}^2M_{u}^2}
\bigg[ \bar{C}_{1}(\tilde{\varsigma}) + 2(\upsilon-1) \nonumber\\
& \quad \times (x \check x m_{K}^2+M_{s}M_{u}+\upsilon k_{\perp}^2)\frac{\bar{C}_{2}(\tilde{\varsigma})}{\tilde{\varsigma}}\bigg]\,,
\end{align}
\end{subequations}
where $\tilde\varsigma =
\upsilon(1-\upsilon) k_{\perp}^{2} + (1 - \upsilon)\zeta_0+\upsilon m_{G}^2$.
%$ \upsilon(1-\upsilon) k_{\perp}^{2} + (1 - \upsilon)[M^2 - x\check x m_{K}^2]+\upsilon m_{G}^2$.
%
Now, alike with QCD and in contrast to Eq.\,\eqref{KaonMIBMFFX},
$h_{1K}^{u\perp} (x, k_{\perp}^{2}) \sim f_{1K}^u(x,k_\perp^2)/k_\perp^2$ at ultraviolet momenta.

It only remains to choose a value for the coupling, $\alpha_{\mathpzc L}$, in Eq.\,\eqref{momentumglue}.
Following the analogous pion study \cite{Cheng:2024gyv}, we use
\begin{equation}
\alpha_{\mathpzc L} = 0.97 \pi \,,
\label{alphaL}
\end{equation}
\emph{i.e}.\ the infrared value of QCD's process-independent effective charge \cite{Cui:2019dwv, Deur:2023dzc, Brodsky:2024zev}.
This value is known with a precision of 4\%.

\begin{figure}[t]
\vspace*{0.4em}

\leftline{\hspace*{0.5em}{\large{\textsf{A}}}}
\vspace*{-2ex}
\includegraphics[width=0.41\textwidth]{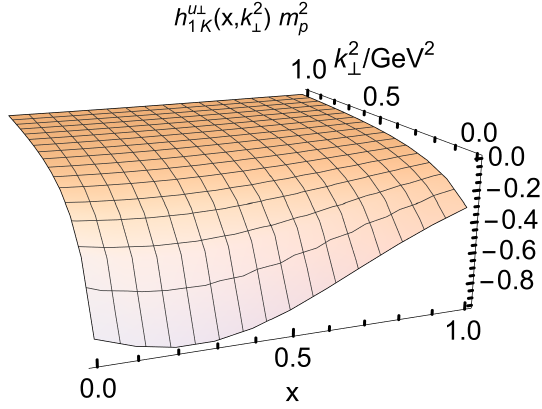}
\vspace*{0.1ex}
\leftline{\hspace*{0.5em}{\large{\textsf{B}}}}
\vspace*{-2ex}
\includegraphics[width=0.41\textwidth]{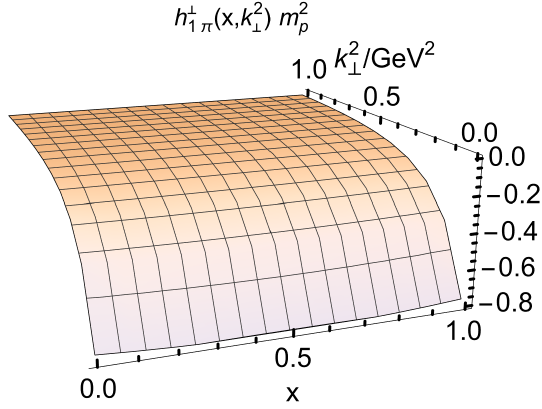}
\vspace*{3.5ex}
\caption{\label{Ploth1q2}
{\sf Panel A}. Hadron scale SCI result for the $u$-in-$K$ BM function drawn from Eqs.\,\eqref{KaonMDBMFFX}, \eqref{alphaL}.  The $s$-in-$K$ BM function is obtained via Eq.\,\eqref{BMSymmetry}.
{\sf Panel B}. Analogous $u$-in-$\pi$ BM function.}
\end{figure}

The BM function obtained from Eqs.\,\eqref{KaonMDBMFFX}, \eqref{alphaL} is drawn in Fig.\,\ref{Ploth1q2}\,A.
It is worth reiterating that this is the result at the hadron scale, $\zeta_{\cal H}$, whereat valence dof carry all hadron properties \cite{Yin:2023dbw}.
Compared with $f_{1K}^u(x,k_\perp^2)$ in Fig.\,\ref{Plotf1}, the BM function has a similar (albeit opposite sign) profile; however, the additional $1/k_\perp^2$ factor is evident.
Comparing with $h_{1\pi}^{\perp}$, Fig.\,\ref{Ploth1q2}\,B, the remarks made in the two paragraphs following Eq.\,\eqref{HIkaonTMDFX} are also pertinent here.

\begin{figure}[t]
\vspace*{0.4em}

\leftline{\hspace*{0.5em}{\large{\textsf{A}}}}
\vspace*{-2ex}
\includegraphics[width=0.41\textwidth]{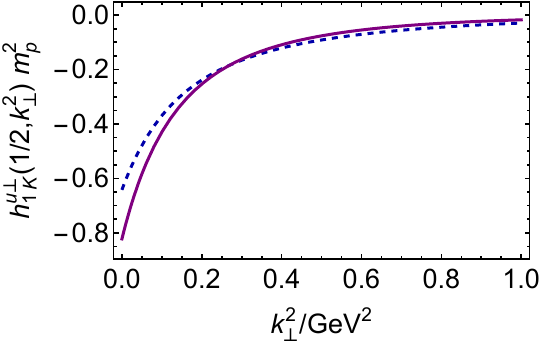}
\vspace*{0.1ex}
\leftline{\hspace*{0.5em}{\large{\textsf{B}}}}
\vspace*{-2ex}
\includegraphics[width=0.41\textwidth]{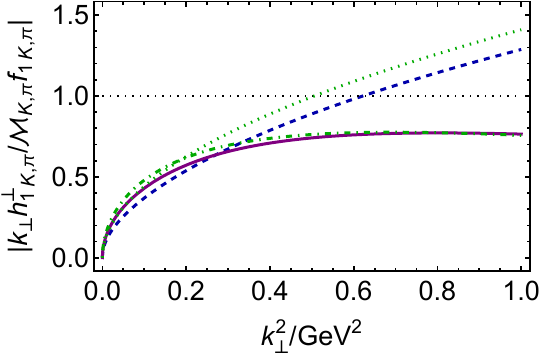}
\vspace*{3.5ex}
\caption{\label{Ploth1pb}
{\sf Panel A}.
$(x=1/2,k_\perp^2)$-dependence of kaon BM functions:
solid purple curve -- Eq.\,\eqref{KaonMDBMFFX}, momentum-dependent gauge link completion;
and dashed blue curve -- Eq.\,\eqref{KaonMIBMFFX}, pure-SCI result.
{\sf Panel B}.
Checking the positivity bound, Eq.\,\eqref{EqPositive}: the bound is violated by any result that crosses the horizontal dotted line.
Legend:
green curves, SCI pion results -- dotted, momentum-independent gluon; dot-dashed, momentum-dependent gauge link completion; otherwise as in {\sf Panel A}.
%
%$|k_\perp h_{1K}^{u\perp {\rm MI}}(x=1/2,k_\perp^2)/M_u f_{1K}^{u}(x=1/2,k_\perp^2)|$ -- dashed blue \emph{cf}.\ $|k_\perp h_{1K}^{u\perp}(x=1/2,k_\perp^2)/M_{u} f_{1K}^{u}(x=1/2,k_\perp^2)|$ solid purple.
}
\end{figure}

\subsection{Positivity bound}
\label{SecPositivity}
We now return to the positivity bound defined by Eq.\,\eqref{EqPositive}.
Like the pion study in Ref.\,\cite{Cheng:2024gyv}, we have calculated the kaon BM function using two gauge link completions: Eqs.\,\eqref{SCIglue}, \eqref{momentumglue}.
Consider, therefore, Fig.\,\ref{Ploth1pb}.
Panel A depicts the $k_\perp^2$-dependence of $k_\perp h_{1K}^{u\perp}(x=1/2,k_\perp^2)$.  Naturally, both completions deliver a negative-definite result; however, the momentum-dependent gluon propagator delivers a BM function that decreases more rapidly with increasing $k_\perp^2$, \emph{viz}.\ it is a softer function of $k_\perp^2$.

Figure~\ref{Ploth1pb}\,B displays the $u$-in-$K$ positivity bound \linebreak curve.
Plainly, whilst the pure-SCI gauge-link completion delivers a BM function that violates Eq.\,\eqref{EqPositive}, the momentum-dependent gluon completion is compatible with the bound.
In the latter case, the good outcome is obtained because the employed quark and gluon propagators possess an ultraviolet momentum dependence that matches QCD expectations, up to logarithmic scaling violations, and the support of the usual quark loop is not curtailed without at least commensurate and consistent suppression of the gauge link range.
(Owing to Eq.\,\eqref{BMSymmetry}, the $s$-in-$K$ results are qualitatively identical.)

In addition to the kaon results, Fig.\,\ref{Ploth1pb}\,B also displays the pion positivity-bound curve obtained with Eqs\,\eqref{SCIglue} [dotted green] and \eqref{momentumglue} [dot-dashed green].
Evidently, as noted previously \cite{Cheng:2024gyv}, the curve increases more slowly when the bound-state mass is larger.

\section{Calculation using Light-Front Wave Function}
\label{LFWFcalc}
Supposing one had developed a light-front QCD Hamiltonian, this could then be used to define a Fock space expansion in which the leading two-particle element for the kaon corresponds to the rainbow-ladder (RL) truncation bound state discussed, \emph{e.g}., in \ref{AppendixSCI}.
At this point, the LFWF associated with the leading term in the operator expansion of the Fock space corresponds to the kaon's RL LFWF.
Focusing on the $u$-in$K$ component, we write this as
\begin{equation}
\Psi_K^u(x,k_\perp;\lambda_1,\lambda_2)
=\psi_K^u(x,k_\perp^2) S^{u_K}_{\lambda_1,\lambda_2}(x,\vec{k}_{\perp})\,,
\label{kaonLFWF}
\end{equation}
where the second term records the helicities of the identified quasiparticles.
(Such an assumption is implicit in every light-front model of hadron structure.  Further, at $\zeta_{\cal H}$,
$\Psi_K^s(x,k_\perp;\lambda_1,\lambda_2) = \Psi_K^u(1-x,k_\perp;\lambda_1,\lambda_2)$.)
Equation~\eqref{kaonLFWF} is just the LFWF that is obtained when one completes a light-front projection of the Poincar\'e-co\-va\-riant RL-truncation kaon Bethe-Salpeter wave function.

Working as just described, one finds
\begin{subequations}
\label{LFWF}
\begin{align}
\psi_K^u(x,k_\perp^2) & =
\frac{  \sqrt{2 N_c} /[x \check x]}{[(k_\perp^2+M_u^2)/x+(k_\perp^2+M_s^2-\check x m_K^2)/\check x]}\\
% (k_T^2+M_u^2)/x+(k_T^2-\check{x}m_K^2+M_s^2)/(1-x)
%
& = \frac{\sqrt{2N_c}}{k_{\perp}^2+xM_{s}^{2} + \check x M_{u}^{2} - x \check x m_K^2} \,,
\end{align}
\end{subequations}
and
\begin{align}
\label{SSKLFWF}
S_{\lambda_1,\lambda_2}^{u_K}(x,\vec k_\perp)  & =
\frac{1}{2M_{su}}\left[
\begin{array}{cc}
S_{\uparrow\uparrow} & S_{\uparrow\downarrow}\\
S_{\downarrow\uparrow} & S_{\downarrow\downarrow}
\end{array}
\right]\,,
\end{align}
with $M_{su}=M_s M_u/(M_s+M_u)$,
\begin{subequations}
\begin{align}
S_{\uparrow\uparrow} & = -\left[2E_{K} M_{su}-F_{K}(M_s+M_u)\right](k_{1}-ik_{2})\,, \\
S_{\uparrow\downarrow} & = 2 E_{K} M_{su} \left[\check x M_u + xM_s \right]\nonumber \\
& \quad + F_{K} [(k_\perp^2-M_s M_u) - x\check xm_{K}^2]\,,
\end{align}
\end{subequations}
$S_{\downarrow\downarrow} = S_{\uparrow\uparrow}^\ast $,
$S_{\downarrow\uparrow} = - S_{\uparrow\downarrow} $.
Taking $s\to d$, one recovers the analogous pion LFWF; see Ref.\,\cite[Eq.\,(19)]{Cheng:2024gyv}.

Working with the LFWF, the helicity-independent TMD is:
\begin{align}
\label{UnpolarizedTMD_LFWF}
f_{1K}^u(x,\vec{k}_{\perp})
& = \tfrac{1}{16\pi^3}\sum_{\lambda_{1}\lambda_{2}} |\psi_K^u \left(x, k_{\perp}^2\right)|^2 \nonumber \\
& \qquad \times S_{\lambda_{1},\lambda_{2}}^{{u_K} \dagger}(x,\vec{k}_{\perp}) S_{\lambda_{1},\lambda_{2}}^{u_K}(x,\vec{k}_{\perp})\,.
\end{align}
Obviously, as the modulus-squared of the LFWF, this is a positive definite quantity; so, can properly be interpreted as a probability density.
Moreover,  from this point, straightforward algebra reproduces Eq.\,\eqref{uinKTMD}.

The analogous expression for the BM function is:
\begin{subequations}
\label{LFWFBM}
\begin{align}
\label{BM_LFWF}
\tfrac{k_\perp^2}{{\mathpzc M}_K} & h_{1K}^{u \perp}(x,\vec{k}_{\perp})  =
\int\frac{d^2k_{\perp}^\prime}{16\pi^3} \mathcal{G}(x,\vec{k}_{\perp},\vec{k}_{\perp}^\prime)
\nonumber \\
& \qquad \times \psi_K^{u\ast}(x,k_{\perp}^\prime) \psi_K^u(x,k_{\perp})
i\Theta(x,\vec{k}_{\perp},\vec{k}_{\perp}^\prime)\,, \\
\Theta&(x,\vec{k}_{\perp},\vec{k}_{\perp}^\prime)
= \sum_{\lambda_1,\lambda_2} S^{{u_K}\dagger}_{-\lambda_1,\lambda_2}(x,\vec{k}_{\perp}^\prime)
\nonumber \\ %
& \qquad\times \lambda_1k_\perp e^{i\lambda_1\theta_{k_\perp}}S_{\lambda_1,\lambda_2}^{u_K}(x,\vec{k}_{\perp}),
\end{align}
\end{subequations}
where $\theta_{k_{\perp}}$ is the angle between $O^{\perp}_{\alpha\mu} k_\mu$ and $O^{\perp}_{\alpha\mu} k_\mu^\prime$ in the two-dimensional plane they define.
In this expression, $\mathcal{G}(x,\vec{k}_{\perp},\vec{k}_{\perp}^\prime)$ encodes the gauge link contribution and takes the form \cite{Lu:2006kt}:
\begin{equation}
{\cal G}(x,\vec{k}_\perp,\vec{k}_\perp^\prime)
= \frac{i \alpha}{2\pi}  D(q_\perp^2)\,,
\end{equation}
with $q_\perp = k_\perp - k_\perp^\prime$ and $D(q_\perp^2)$ is a function to be specified.   Using Eq.\,\eqref{SCIglue} or \eqref{momentumglue}, one recovers either Eq.\,\eqref{KaonMIBMFFX} or \eqref{KaonMDBMFFX}.  These outcomes confirm that one has complete consistency between the diagrammatic and LFWF approaches.

\section{TMD Evolution}
\label{TMDEvolve}
As noted in Sect.\,\ref{sec1}, TMD scale evolution is more complicated than that for collinear DFs.  Given the simplicity of the SCI, we elect to avoid the associated complications by working with TMD moments.
\emph{N.B}.\ All results in this section are calculated using TMDs obtained with the momentum-dependent gauge link completion, Eq.\,\eqref{momentumglue}.

\subsection{Leading $k_\perp^2$ moment}
Consider the following $k_\perp^2$-moment of the hadron-scale BM function:
\begin{subequations}
\label{k2mom1}
\begin{align}
 h_{1K}^{q\perp(1)}(x) & =\frac{1}{2 {\mathpzc M}_K^2}\int d^2 \vec{k}_\perp k_\perp^2 h_{1K}^{q\perp}(x,k_\perp^2) \\
& = \tfrac{1}{2 \mathcal M_K} T_{uK}^{(\sigma)}(x,x)\,,
\label{kperpmom}
\end{align}
\end{subequations}
where $T_{uK}^{(\sigma)}$ is a twist-three quark+gluon+quark correlation function \cite{Kang:2012em}.  The scale evolution equation for this function is known \cite{Kang:2012em}:
\begin{align}
& \frac{\partial T_{q,F}^{(\sigma)}(x,x,\zeta)}{\partial\ln{\zeta^2}}  = \frac{\alpha_s(\zeta^2)}{2\pi}\int_x^1\frac{d\xi}{\xi}\left\{\Delta_T P_{qq}(z)T_{q,F}^{(\sigma)}(\xi,\xi,\zeta)\nonumber\right.\\
&\qquad \left. + \frac{N_c}{2} \left[\frac{2T_{q,F}^{(\sigma)}(\xi,x,\zeta) - 2zT_{q,F}^{(\sigma)}(\xi,\xi,\zeta)}{1-z}\right]\nonumber\right.\\
&\qquad \left.-N_c\delta(1-z)T_{q,F}^{(\sigma)}(x,x,\zeta)\nonumber\right.\\
&\qquad \left.+\frac{1}{2N_c}\left[2(1-z)T_{q,F}^{(\sigma)}(x,x-\xi,\zeta)\right]\right\}\,,
\label{CCEVE}
\end{align}
where $\Delta_T P_{qq}(z=x/\xi)$ is the transversity splitting kernel
\begin{align}
\Delta_T P_{qq}(z)=C_F\left[\frac{2z}{(1-z)_+}+\frac{3}{2}\delta(1-z)\right]\,,
\label{CCEVKer}
\end{align}
$C_F = (N_c^2 - 1)/(2 N_c)$, and the definition of ``$1/(1-z)^+$'' is standard -- see, \emph{e.g}., Ref.\,\cite[Eq.\,(5.34)]{Roberts:1990ww}.

The dependence of the Eq.\,\eqref{CCEVE} integrand on\linebreak $T_{q,F}^{(\sigma)}(\xi,x\neq\xi,\zeta)$, \emph{i.e}., its nondiagonal character, is a complicating feature of this evolution equation.
Hitherto, practitioners have typically neglected the off-dia\-go\-nal contributions, truncating the kernel and keeping only the first line and, sometimes, also the third line \cite{Cheng:2024gyv, Wang:2017onm, Kou:2023ady, Echevarria:2020hpy}.  Herein, we move beyond that truncation in a manner which enables the impact of the off-diagonal terms to be illustrated, if not rigorously controlled.  To that end, we follow Ref.\,\cite{Kang:2008ey} and write \begin{align}
T_{q,F}^{(\sigma)}&(x_1,x_2,\zeta) = e^{-\frac{(x_1-x_2)^2}{2\sigma^2}}\nonumber \\
& \times \tfrac{1}{2}\left[T_{q,F}^{(\sigma)}(x_1,x_1,\zeta)+T_{q,F}^{(\sigma)}(x_2,x_2,\zeta)\right]
\,,
\label{CCModel}
\end{align}
where $\sigma$ is a (dimensionless) width parameter that characterises the off-diagonal persistence of the correlation.

In proceeding, we interpret Eq.\,\eqref{CCEVKer} via the all-orders (AO) evolution scheme described in Ref.\,\cite{Yin:2023dbw}.
This approach has proved efficacious in numerous applications, \emph{e.g}.,
delivering unified predictions for all pion, kaon, and proton (unpolarised and polarised) DFs \cite{Cui:2020tdf, Chang:2022jri, Lu:2022cjx, Cheng:2023kmt, Yu:2024qsd}, that agree with much available data;
pion and kaon fragmentation functions \cite{Xing:2023pms, Xing:2025eip};
and a species decomposition of nucleon gravitational form factors \cite{Yao:2024ixu}.
In the AO scheme, $\alpha(\zeta^2)$ is an effective charge \cite{Grunberg:1980ja, Grunberg:1982fw, Deur:2023dzc}, namely, a QCD running coupling that is defined so that, when used to integrate the leading-order perturbative DGLAP equations \cite{Dokshitzer:1977sg, Gribov:1971zn, Lipatov:1974qm, Altarelli:1977zs}, it delivers an evolution scheme for all DFs -- unpolarised and polarised, and for any hadron -- that is all-orders exact.
Defined in this way, $\alpha(\zeta^2)$ implicitly incorporates terms of arbitrarily high order in the perturbative coupling.  Any such effective charge has numerous valuable qualities, \emph{e.g}., it is: consistent with the QCD renormalisation group; independent of the renormalisation scheme; analytic and finite; and provides an infrared completion of any standard running coupling.

\begin{table}[t]
\centering
\caption{Dependence of
${\mathsf h}_{K,\pi}^{q(01)}(\zeta_2=2\,{\rm GeV}) /{\mathsf h}_{K,\pi}^{q(01)}(\zeta_{\cal H}) $, Eq.\,\eqref{ZeroH}, on the choice of evolution kernel in Eq.\,\eqref{CCEVE}.
Of course, ${\mathsf h}_{K}^{u(01)}(\zeta_{\cal H})={\mathsf h}_{K}^{s(01)}(\zeta_{\cal H})$, with ${\mathsf h}_{K}^{u(01)}(\zeta_{\cal H})= - 0.38$, and
${\mathsf h}_{\pi}^{\ell(01)}(\zeta_{\cal H})= -0.42$.
\label{h0moment}}
\begin{tabular*}
{\hsize}
{
l@{\extracolsep{0ptplus1fil}}
l@{\extracolsep{0ptplus1fil}}
|l@{\extracolsep{0ptplus1fil}}
l@{\extracolsep{0ptplus1fil}}
l@{\extracolsep{0ptplus1fil}}}\hline
kernel $[P_{qq}(z)]\ $ & $\sigma\ $ & $u$-in$K\ $&  $\bar s$-in$K\ $ & $u=\bar d$-in $ \pi\  $ \\\hline
%% & Total & $u\ $ & $d\ $ & $s\ $ & $c\ $ & $q\ $ & $g\ $ \\\hline
$\Delta_T P_{qq}(z)\ $ Eq.\,\eqref{CCEVKer}
&  & $0.76\ $ & $0.76\ $ & $0.76\ $ \\
$\Delta_T P_{qq}(z) - N_c\delta(1-z)\ $ &  & $0.22\ $ & $0.22\ $ & $0.22\ $ \\ \hline
Off-diagonal Eq.\,\eqref{CCModel}
& $1.0\ $ & $2.50\ $ & $1.95\ $ & $2.34\ $ \\
& $0.5\ $ & $1.89\ $ & $1.46\ $ & $1.76\ $ \\
& $0.25\ $ & $1.14\ $ & $0.85\ $ & $1.06\ $ \\
& $0.17\ $ & $0.82\ $ & $0.60\ $ & $0.76\ $ \\
& $0.044\ $ & $0.24\ $ & $0.16\ $ & $0.22\ $ \\
\hline
\end{tabular*}
\end{table}

Consider now the following zeroth $x$-moment:
\begin{equation}
{\mathsf h}_{K}^{q\perp(01)}(\zeta) = \int_0^1 dx\,h_{1K}^{q\perp(1)}(x;\zeta) \,.
\label{ZeroH}
\end{equation}
Owing to the presence of the off-diagonal terms, the simplest way to calculate the evolution of this moment is to first solve the $x$-dependent problem and then integrate.  Notwithstanding, this one quantity is a useful indicator of the $\sigma$-dependence of evolution, with the results displayed in Table~\ref{h0moment}.

The first observation is that with the simple kernels, which involve no off-diagonal contributions, the ratio depends on the kernel choice but is independent of the input function.  This feature is readily understood because, in the AO scheme, as at leading order in perturbation theory:
\begin{align}
    {\mathsf h}_{K,\pi}^{q\perp(01)}(\zeta_2) & /{\mathsf h}_{K,\pi}^{q\perp(01)}(\zeta_{\cal H})  \nonumber \\
       & =
    \exp\left[ \gamma_h^0 \int_{\zeta_1^2}^{\zeta_2^2} \!ds \,\alpha(s)/(2\pi s)]
    \right]\,.
    \label{EvolutionHn1}
\end{align}
The AO QCD effective charge, $\alpha(s)$, mat\-ches the perturbative charge on $s\gtrsim 2 m_p^2$ \cite[Fig.\,3]{Ding:2022ows}, and, in this $\sigma$-independent case, the anomalous dimensions are computed from the simple splitting function employed in the kernel, which is listed in the first column of Table~\ref{h0moment}\,-\,Rows~1, 2:
\begin{equation}
\gamma_h^m  = \int_0^1 \! dz \, z^m P_{qq}(z) \,.
\end{equation}

\begin{figure}[t]
\vspace*{0.4em}

\leftline{\hspace*{0.5em}{\large{\textsf{A}}}}
\vspace*{-2ex}
\includegraphics[width=0.43\textwidth]{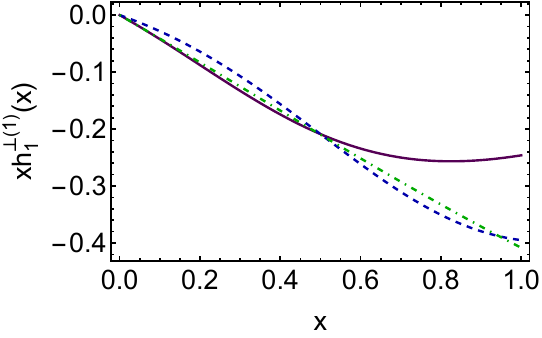}
\vspace*{0.1ex}
\leftline{\hspace*{0.5em}{\large{\textsf{B}}}}
\vspace*{-2ex}
\includegraphics[width=0.43\textwidth]{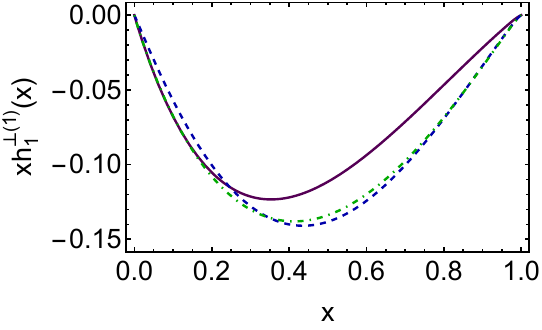}
\vspace*{0.1ex}
\leftline{\hspace*{0.5em}{\large{\textsf{C}}}}
\vspace*{-2ex}
\includegraphics[width=0.44\textwidth]{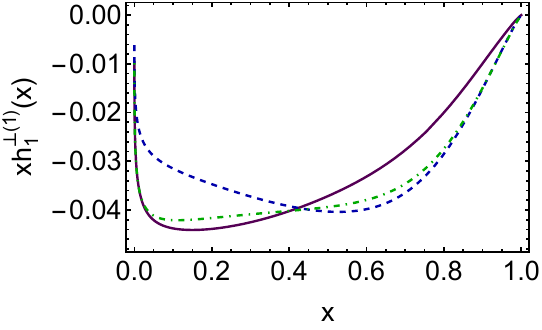}
\vspace*{3.5ex}

\caption{\label{Figh10}
$x {h}_{K,\pi}^{q\perp(01)}(x;\zeta)$.
{\sf Panel A}.  Hadron scale.
{\sf Panel B}.  After evolution $\zeta_{\cal H} \to \zeta_2$ using a simple evolution kernel -- Eq.\,\eqref{CCEVKer} only.
{\sf Panel C}.  After evolution $\zeta_{\cal H} \to \zeta_2$ using the off-diagonal evolution kernel -- Eqs.\,\eqref{CCEVE}, \eqref{CCModel} with $\sigma = 0.17$.
Legend.
$x { h}_{K}^{u\perp(01)}(x;\zeta)$ -- solid purple curve;
$x { h}_{K}^{s\perp(01)}(x;\zeta)$ -- dashed blue;
$x { h}_{\pi}^{u\perp(01)}(x;\zeta)$ -- dot-dashed green.  }
\end{figure}

After inclusion of the off-diagonal contributions, the picture is different.
For the pion, with it's symmetric BM function, one sees an increase in
${\mathsf h}_{K,\pi}^{q\perp(01)}(\zeta_2)$ for $\sigma \gtrsim 1/4$ and a decrease otherwise.
Furthermore, with $\sigma = 0.17$, one recovers the simplest-kernel result; $\sigma = 0.044$ delivers the other simple kernel result; see Table~\ref{h0moment}\,-\,Rows~6, 7.
It is worth noting that, after evolution to $\zeta_2$, the $\bar s$-in-$K^+/u$-in-$K^+$ ratio is roughly $5/4$ and fairly insensitive to the value of $\sigma$ in Eq.\,\eqref{CCModel}.

One also notes that the kaon moment ratios become asymmetric under off-diagonal evolution.
This is because
(\emph{a}) the Eq.\,\eqref{CCModel} \emph{Ansatz} for the off-diagonal behaviour disfavours support on $\xi > x$;
(\emph{b}) as readily inferred from Fig.\,\ref{Figh10}\,A, the greater part of the  $h_{1K}^{s\perp(1)}(x;\zeta_{\cal H})$ support lies on $x>1/2$;
(\emph{c}) consequently, the peak magnitude of $h_{1K}^{s\perp(1)}(x;\zeta)$ becomes suppressed with respect to that of $h_{1K}^{u\perp(1)}(x;\zeta)$; see Fig.\,\ref{Figh10}\,C.
It is worth noting that since contemporary predictions for $u$ and $d$ valence quark DFs in the proton also show a relative shift in the peak locations, with that of the $u$ quark lying at larger $x$ -- see Ref.\,\cite[Fig.\,1]{Chang:2022jri}, then one may expect an analogous, albeit more modest, effect for the proton BM functions.

\subsection{BM Shift}
\label{TMDEvolveBMshift}
{\allowdisplaybreaks
Consider the following dimensionless momentum-con\-ju\-gate quantity:
\begin{align}
{\mathpzc f}_{K,\pi}^{[j](n)}(b_\perp^2;\zeta) & =
\frac{n!}{m_{K,\pi}^{2n}}
\int_0^1 \! dx \, x^j \int d k_\perp k_\perp  \nonumber \\
& \times \quad  \left[\frac{k_\perp}{b_\perp }\right]^n
J_n(b_\perp k_\perp){\mathpzc  f_{K,\pi}}(x,k_\perp^2)\,,
\end{align}
where ${\mathpzc f}_{K,\pi}(x,k_\perp^2)$ is some TMD,
$J_n$ is a Bessel function of the first kind,
and $\zeta$ is the resolving scale.
Setting $n=1$, one obtains the so-called generalised BM shift:
\begin{equation}
\langle |k_\perp| \rangle_{\rm UT}(b_\perp^2,\zeta)
= m_{K,\pi}
\frac{\tilde h_{1(K,\pi)}^{\perp [0] (1)}(b_\perp^2;\zeta)}{f_{1(K,\pi)}^{[0],(0)}(b_\perp^2;\zeta)}\,,
\label{GBMshift}
\end{equation}
(unpolarised target ``U'' containing transversely polari\-sed valence dof ``T'')
which, for a heavy pion, $m_\pi = 0.518\,$GeV, has been computed at $\zeta = \zeta_2$ using lattice-regularised QCD (lQCD) \cite{Engelhardt:2015xja}.
In this section
\begin{equation}
\tilde h_{1(K,\pi)}^{\perp} = (m_{K,\pi}/{\mathpzc M}_{K,\pi}) h_{1(K,\pi)}^{\perp}\,,
\label{GenBMShift}
\end{equation}
\emph{i.e}., we rescale our BM function by  $(m_{K,\pi}/{\mathpzc M}_{K,\pi})$ so as to match the convention used in the lQCD study.
}

In the limit $b_\perp^2 \to 0$, the ratio in Eq.\,\eqref{GBMshift} becomes the standard BM shift, \emph{viz}.\ the mean transverse $y$-direc\-tion momentum of $x$-direction polarised quarks in the unpolarisable pseudoscalar meson.  Naturally, Eq.\,\eqref{BNC} entails:
\begin{equation}
f_{1(K,\pi)}^{[0],(0)}(b_\perp^2;\zeta) = 1.
\end{equation}

Working in the neighbourhood $b_\perp^2 \simeq 0$, whereupon
$J_1( b_\perp k_\perp) \approx b_\perp k_\perp/ 2$, one has a mass-rescaled analogue of Eq.\,\eqref{k2mom1}, \emph{viz}.\
\begin{equation}
\tilde h_{1(K,\pi)}^{q\perp(1)}(x)  =\frac{1}{2 m_{K,\pi}^2}\int d^2 \vec{k}_\perp
k_\perp^2 \tilde h_{1(K,\pi)}^{q\perp}(x,k_\perp^2) \,.
\label{NearMom}
\end{equation}
The evolution of this (dimensionless) function is also prescribed by Eq.\,\eqref{CCEVE}; so, one can readily calculate the generalised BM shift at any desired scale. %; see Fig.\,\ref{PlotGshift}.
\emph{N.B}.\ Integrating the function in Eq.\,\eqref{NearMom} over $x$, one recovers the moment defined in Eq.\,\eqref{ZeroH}, up to an overall (dimensionless) constant $= 1/2$.  The implicit presence of $J_1(b_\perp k_\perp)$ in Eq.\,\eqref{GBMshift} breaks this exact correspondence, nevertheless, one should expect good likenesses, \emph{e.g}., a hadron scale BM shift on the order of $-0.2$; see caption of Table~\ref{h0moment}.

\begin{figure}[t]
\centerline{%
\includegraphics[clip, width=0.44\textwidth]{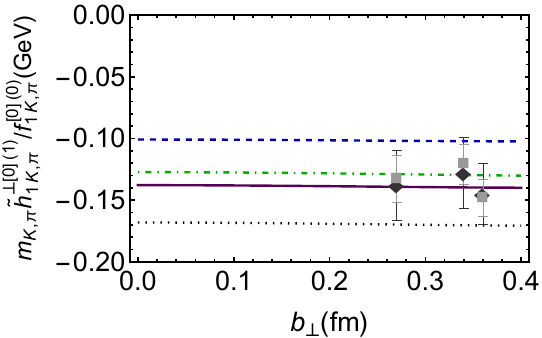}}
\caption{\label{PlotGshift}
Generalised BM shift, Eq.\,\eqref{GenBMShift}.
Legend.
Hadron-scale values for $\pi$, $u$-in-$K$, $s$-in-$K$ -- dotted black curve.  (As explained after Eq.\,\eqref{NearMom}, they are approximately the same, \emph{i.e}., $\approx -0.17$.)
After evolution $\zeta_{\cal H} \to \zeta_2$:
$u$-in-$K$ -- solid purple curve;
$s$-in-$K$ -- dashed blue curve;
$u=d$-in-$\pi$ -- dot-dashed green curve;
points -- available lQCD results \cite{Engelhardt:2015xja}, computed with $m_\pi = 0.518\,$GeV at a resolving scale $\zeta=\zeta_2=2\,$GeV.
\emph{N.B}.\
%In this image, $M$ is the relevant meson mass; and
$0.2\,$GeV$\approx 1 {\rm fm}^{-1}$.}
\end{figure}

The generalised BM shift for kaons and pion is depicted in Fig.\,\ref{PlotGshift}.
Evidently, it is practically $b_\perp^2$-indepen\-dent on the depicted domain; hence, the standard BM shift is readily determined.
As telegraphed following Eq.\,\eqref{NearMom}, at $\zeta_{\cal H}$, with internally consistent normalisation, the dimensionless shift is approximately the same for all pseudoscalar mesons, taking the value \linebreak
$\langle |k_\perp| \rangle_{\rm UT}(b_\perp^2 \to 0 ,\zeta_{\cal H}) = -0.17\,$GeV.
Using $\sigma = 0.17$ in the off-diagonal kernel, which reproduces the Eq.\,\eqref{CCEVKer} diagonal kernel $\pi$ result, one finds that the shift diminishes with increasing scale and it is larger in magnitude for $u$-in-$K$ than $s$-in-$K$:
$\langle |k_\perp| \rangle_{\rm UT}^{u,K}(b_\perp^2 \to 0 ,\zeta_{\cal H}) = -0.14\,$GeV \emph{cf}.\,
$\langle |k_\perp| \rangle_{\rm UT}^{s,K}(0 ,\zeta_{\cal H}) = -0.10$\,GeV.
The pion result lies between these two curves:
$\langle |k_\perp| \rangle_{\rm UT}^{u=d, \pi}( 0 ,\zeta_{\cal H}) = -0.13\,$GeV.
For $\sigma \gtrsim 1/4$, evolution enhances the shift instead, but the relative ordering remains the same.
Evidently, concerning the BM shift, lQCD results with improved precision are necessary before that approach can yield meaningful conclusions about flavour separation in kaon-like systems.

%%u-in-Kaon: -0.1392 (b_T=0.27), -0.1396 (b_T=0.34) and -0.1397 (b_T=0.36) at 2GeV scale.
%%s-in-Kaon:  -0.1022 (b_T=0.27), -0.1022 (b_T=0.34) and -0.1022 (b_T=0.36) at 2GeV scale.
%%pion: -0.1294 (b_T=0.27), -0.1294 (b_T=0.34) and -0.1295 (b_T=0.36) at 2GeV scale.

\section{Summary and Perspective}
\label{epilogue}
A symmetry preserving treatment of a vector\,$\otimes$\,vec\-tor contact interaction (SCI) was used as the basis for delivering predictions for the four ($u$, $s$) nonzero kaon transverse momentum dependent parton distribution functions (TMDs), namely, those for unpolarised valence degrees-of-freedom (dof) and the Boer-Mulders (BM) functions, which describe correlations between valence dof transverse spins and transverse momentum [Fig.\,\ref{FBM}].
Working with the SCI, all analyses are largely algebraic, so the formulae and results exhibit a high level of transparency.  This enables clear insights to be drawn; not just from and about the SCI results themselves, but also regarding those obtained using more sophisticated frameworks through relevant comparisons.  Furthermore, interpreted carefully, SCI results can be physically relevant.
Of particular interest herein are
the comparisons drawn between flavour-separated kaon TMDs themselves;
subsequent comparisons with those of the pion;
and an exposition of the impacts of off-diagonal terms in the evolution kernel for the BM function.

Unpolarised TMDs are always nonzero; and our SCI analysis predicts that, unlike the analogous pion TMD, that for the kaon is asymmetric around $x=1/2$ [Fig.\,\ref{Plotf1}].
Both kaon and pion TMDs are dilated as a consequence of emergent hadron mass (EHM) phenomena; however, Higgs modulation of EHM produces a shift in the peak location of the $u$-in-$K^+$ TMD, \emph{viz}.\ $x=0.5 \to 0.3$.  Naturally, at the hadron scale, $\zeta_{\cal H}$, the $\bar s$-in-$K^+$ TMD is obtained from the $u$-in-$K^+$ function by mapping $x\to (1-x)$.

On the other hand, a nonzero BM function is only possible when, in calculating the associated $\gamma (K,\pi)  \to \gamma (K,\pi)$ matrix elements, one includes interactions between the spectator and the struck and, thereafter, highly energetic valence dof.
Such interactions are described by gauge-field links, which, in continuum analyses, are typically introduced via one or another phenomenological model.
Our analysis employed an eikonal approximation to represent quark propagation under the influence of the gauge link [Fig.\,\ref{ImageBM}, Sect.\,\ref{subsecGL}].
It confirms that the magnitude of the BM function increases with the size of the dressed masses of the valence dof; hence, that the strength of spin-momentum correlations is a signal and measure of EHM.
Moreover and importantly, we also verified that the pointwise positivity constraint [Eq.\,\eqref{EqPositive}] can only be satisfied when due attention is given to building consistency between the support domains of the hadron binding interaction and that used to characterise the gauge link.

So as to make connections between various approa\-ches to TMD computation, we also repeated our analyses using kaon light-front wave functions (LFWFs) built so that consistency with the diagrammatic SCI calculations is ensured [Sec.\,\ref{LFWFcalc}].  Amongst other things, this aspect of our study highlighted the rigour and importance of the positivity bound.

We also discussed some features of TMD evolution [Sect.\,\ref{TMDEvolve}], focusing on the leading $k_\perp^2$ moment of the BM function and the related BM shift.
Of particular interest is our analysis of the impacts of off-diagonal terms in the evolution kernel appropriate to these quantities: they are significant and, therefore, introduce uncertainty into the predictions.
In addition, the off-diagonal terms lead to a flavour separation between the evolved moments of $u$-in-$K$ and $s$-in-$K$ BM functions.

A useful extension of this study would be to build upon Refs.\,\cite{Yu:2024qsd, Bai:2026nqo} and use the SCI to calculate proton BM functions.  Such analyses have the potential to reveal novel impacts of diquark correlations on nucleon structure.
Furthermore, one could calculate the TMDs discussed herein using realistic, QCD-connected pion and kaon LFWFs, which are now available \cite{Yao:2025xjx, Xiao:2025cqz}.
Both efforts are underway.

%\noindent\textbf{Acknowledgments}.
\begin{acknowledgements}
D.-D.\ Cheng is grateful for the hospitality of ECT* members during a one-year visit, which saw part of this work completed, and acknowledges useful communications with H.-Y.\ Xing.
%
%We are grateful for useful discussions with J.\,Rodr\'{\i}guez-Quintero, C.\,Shi and H.-Y.\,Xing.
%
Work supported by:
National Natural Science Foundation of China grant no.\,12135007;
and
China Scholarship Council grant no.\ 202406190234.

\medskip

\noindent\textbf{Data Availability Statement} Data will be made available on reasonable request.  [Authors' comment: All information necessary to reproduce the results described herein is contained in the material presented above.]
\medskip

\noindent\textbf{Code Availability Statement} Code/software will be made available
on reasonable request. [Authors' comment: No additional remarks.]

\end{acknowledgements}

\appendix

\section{SCI}
\label{AppendixSCI}
\subsection{Special functions}
Functions of the following type arise in SCI bound-state equations:
%\begin{subequations}
\begin{align}
%\overline{\cal C}^{\rm iu}_0(\sigma) & =\Gamma(-1,\sigma \tau_{\textrm{uv}}^{2}) - \Gamma(-1,\sigma \tau_{\textrm{ir}}^{2}), \\
%
%\overline{\cal C}^{\rm iu}_1(\sigma)& = \Gamma(0,\sigma \tau_{\textrm{uv}}^{2}) - \Gamma(0,\sigma \tau_{\textrm{ir}}^{2}), \\
%
%\overline{\cal C}^{\rm iu}_2(\sigma) & = \Gamma(1,\sigma \tau_{\textrm{uv}}^{2}) - \Gamma(1,\sigma \tau_{\textrm{ir}}^{2})\,,\\
%
n !\, \overline{\cal C}^{\rm iu}_n(\sigma) & = \Gamma(n-1,\sigma \tau_{\textrm{uv}}^{2}) - \Gamma(n-1,\sigma \tau_{\textrm{ir}}^{2})\,,
\label{eq:Cn}
\end{align}
%\end{subequations}
where $\tau_{\rm ir}$, $\tau_{\rm uv}$ are SCI parameters, ${\cal C}^{\rm iu}_n(\sigma)=\sigma \overline{\cal C}^{\rm iu}_n(\sigma)$, $n\in {\mathbb Z}^\geq$, with $\Gamma(x,y)$ being the incomplete gamma function.
In connection with these functions, there is a useful integration rule:
\begin{equation}
\label{IntCbar}
\int_0^\infty dy \, (n+1)! \frac{\overline{\cal C}^{\rm iu}_{n+1}(y+y_0)}{(y+y_0)^{n}}
= n! \frac{\overline{\cal C}^{\rm iu}_n(y_0)}{y_0^{n-1}}\,.
\end{equation}

\subsection{Interaction}
The SCI is described in many sources.
Here, for internal completeness, we reproduce and somewhat augment material from Refs.\,\cite[Appendix~A.2]{Yu:2024qsd}, \cite[Sec.\,2]{Xu:2021iwv}.
As therein, our analysis is performed at leading-order in the systematic, symmetry preserving approximation \linebreak scheme for the continuum bound state problem introduced in Refs.\,\cite{Munczek:1994zz, Bender:1996bb}, \emph{i.e}., rainbow-ladder (RL) truncation.
At this level, the basis for any continuum meson bound-state problem is the quark + antiquark scattering kernel, which can be written:
\begin{align}
\label{KDinteraction}
\mathscr{K}_{\alpha_1\alpha_1',\alpha_2\alpha_2'}  & = \tilde{\mathpzc G}(k^2) T^k_{\mu\nu} [i\gamma_\mu]_{\alpha_1\alpha_1'} [i\gamma_\nu]_{\alpha_2\alpha_2'}\,,
\end{align}
where $k = p_1-p_1^\prime = p_2^\prime -p_2$, with $p_{1,2}$, $p_{1,2}^\prime$ being the initial and final momenta, respectively, of the scatterers, and $k^2T_{\mu\nu}^k = k^2\delta_{\mu\nu} - k_\mu k_\nu$.

In Eq.\,\eqref{KDinteraction}, the key piece is $\tilde{\mathpzc G}$.
Referring to analyses of QCD gauge sector dynamics \cite{Gao:2017uox, Cui:2019dwv}, it is apparent that a gluon mass-scale emerges in QCD \cite{Gao:2017uox, Cui:2019dwv}; hence, $\tilde{\mathpzc G}$ is nonzero and finite at infrared momenta:
\begin{align}
\label{SimpG}
\tilde{\mathpzc G}(k^2) & \stackrel{k^2 \simeq 0}{=} \frac{4\pi \alpha_{\rm IR}}{m_G^2}\,,
\end{align}
with \cite{Cui:2019dwv, Deur:2023dzc, Brodsky:2024zev}: $m_G \approx 0.5\,$GeV, $\alpha_{\rm IR} \approx \pi$.
Herein, we keep the QCD value of $m_G$.
Furthermore, exploiting the fact that a SCI does not support relative momentum between bound-state valence dof, one may simplify the tensor in Eqs.\,\eqref{KDinteraction}:
\begin{align}
\label{KCI}
\mathscr{K}_{\alpha_1\alpha_1',\alpha_2\alpha_2'}^{\rm CI}  & = \frac{4\pi \alpha_{\rm IR}}{m_G^2}
 [i\gamma_\mu]_{\alpha_1\alpha_1'} [i\gamma_\mu]_{\alpha_2\alpha_2'}\,.
 \end{align}

Confinement is introduced by including an infrared mass scale, $\Lambda_{\rm ir}$, when solving all equations related to bound-state problems \cite{Ebert:1996vx}.
This scale ensures the absence of quark + antiquark production thresholds \cite{Krein:1990sf}.
The standard choice is $\Lambda_{\rm ir} = 0.24\,$GeV\,$=1/[0.82\,{\rm fm}]$ \cite{GutierrezGuerrero:2010md}, \emph{i.e}., a confinement length scale that roughly matches the proton size \cite{Cui:2022fyr}.

Of course, SCI integrals also require ultraviolet regularisation.
This destroys the link between ultraviolet and infrared scales that is a distinguishing feature of QCD.  Thus, the associated ultraviolet mass-scale, $\Lambda_{\rm uv}$, becomes a physical parameter, which is fairly interpreted as an upper limit on the domain whereupon amplitudes describing the associated bound-states are practically momentum-independent.

\begin{table}[t]
\caption{\label{Tab:DressedQuarks}
SCI input coupling, $\alpha_{\rm IR}$, ultraviolet cutoff, $\Lambda_{\rm uv}$, and current-quark masses, $m_{u,s}$, that enable a good description of flavour-nonsinglet pseudoscalar meson properties.
As usual, $m_G=0.5\,$GeV, $\Lambda_{\rm ir} = 0.24\,$GeV.
Calculated results: $M_{u,s}$, $m_{\pi,K}$, $f_{\pi,K}$.
With our normalisation, empirical values for the decay constants are $0.092$, $0.11$, respectively.
%
%A value of $f_{\eta_b}=0.47$ is taken from lQCD \cite{McNeile:2012qf}.
%
%
(We assume isospin symmetry and list dimensioned quantities in GeV.)}
\begin{center}
\begin{tabular*}%{|c|c|c|c|c|c|}\hline
{\hsize}
{
%l@{\extracolsep{0ptplus1fil}}|
l@{\extracolsep{0ptplus1fil}}
|c@{\extracolsep{0ptplus1fil}}
c@{\extracolsep{0ptplus1fil}}
c@{\extracolsep{0ptplus1fil}}
c@{\extracolsep{0ptplus1fil}}
|c@{\extracolsep{0ptplus1fil}}
c@{\extracolsep{0ptplus1fil}}
c@{\extracolsep{0ptplus1fil}}
c@{\extracolsep{0ptplus1fil}}
c@{\extracolsep{0ptplus1fil}}
c@{\extracolsep{0ptplus1fil}}}\hline
$\alpha_{\rm IR}\ $ & $\alpha_{\rm IR}/\pi\ $ & $\Lambda_{\rm uv}$ & $m_u$ & $m_s$  &   $M_u$ & $M_s\ $ & $m_{\pi,K}$ & $f_{\pi,K}$ \\\hline
$\ell=u/d\ $  & $0.36\ $ & $0.91\ $ & $0.007\ $ & & $0.37\ $ & &  $0.14\ $ & $0.10\ $  \\\hline
%$\pi\ $  & $l=u/d\ $  & $0.93$ & $0.91\ $ & $0.007\ $ & 0.37$\ $ & 0.14 & 0.10  \\\hline
%$1.03\phantom{2}$ & $0.94\ $ & $0.0068\ $ & $0.16\ $& 0.37$\ $ & $0.53\ $ & 0.50 & 0.11 \   \\\hline
$s$ & $0.33\ $ & $0.94\ $ & & $0.16\ $& & $0.53\ $ & $0.50\ $ & $0.11\ $  \\\hline
%$s$  & $0.93 $ & 0.905 & 0.17 & 0.533 & 0.50 & 0.106 \\\hline
%$K\ $ & $\bar s$  & $0.84$ & $0.94\ $ & $0.16\phantom{7}\ $ & 0.53$\ $ & 0.50 & 0.11 \\\hline
%$K\ $ & $\bar s$  & $0.33\phantom{2}$ & $0.94\ $ & $0.16_s\phantom{7777}\ $ & 0.53$\ $ & 0.50 & 0.11 \\\hline
%$c$  & 0.438 & 1.878 & 1.235 & 1.603 & 3.177 & \\\hline
%$D\ $ & $c$  & $0.32$ & $1.36\ $ & $1.39\phantom{7}\ $ & 1.57$\ $ & 1.87 & 0.15 \\\hline
%$D\ $ & $c$  & $0.12\phantom{2}$ & $1.36\ $ & $1.39_c\phantom{7777}\ $ & 1.57$\ $ & 1.87 & 0.15 \\\hline%$b$  & 0.097 & 3.495 & 4.669 & 4.829 & 3.175 &
%$B\ $ & $\bar b$  & $0.13$ & $1.92\ $ & $4.81\phantom{7}\ $ & 4.81$\ $ & 5.30 & 0.14 \\
%$B\ $ & $\bar b$  & $0.052$ & $1.92\ $ & $4.81_b\phantom{7777}\ $ & 4.81$\ $ & 5.30 & 0.14
%\\\hline
% bbar 0.0116506 Pi
\end{tabular*}
\end{center}
\end{table}

\subsection{Gap equation}
For a quark of flavour $f$, the SCI gap equation is
\begin{align}
\label{GapEqn}
S_f^{-1}(p)  & = i\gamma\cdot p +m_f \nonumber \\
& \quad + \frac{16 \pi}{3} \frac{\alpha_{\rm IR}}{m_G^2}
\int \frac{d^4q}{(2\pi)^4} \gamma_\mu S_f(q) \gamma_\mu\,,
\end{align}
where $m_f$ is the associated quark current-mass, given in Table~\ref{Tab:DressedQuarks} Employing a Poincar\'e-invariant regularisation, the solution takes the form:
\begin{equation}
\label{genS}
S_f^{-1}(p) = i \gamma\cdot p + M_f\,,
\end{equation}
where the SCI dressed-quark's momentum-independent dynamically generated mass, $M_f$  is obtained by solving:
\begin{equation}
%M = m +  \frac{M}{3\pi^2 m_G^2} \,{\cal C}(M^2;\tau_{\rm ir},\tau_{\rm uv})\,,
%M = m +  \frac{M}{3\pi^2 m_G^2} \,{\cal C}^{\rm iu}(M^2)\,,
M_f = m_f + M_f\frac{4\alpha_{\rm IR}}{3\pi m_G^2}\,\,{\cal C}_0^{\rm iu}(M_f^2)\,.
\label{gapactual}
\end{equation}

Herein, we implement SCI regularisation by exploiting a di\-men\-sional-regularisation-like identity:
\begin{equation}
0 = \int_0^1d\alpha \,
%\big[ {\cal C}^{\rm iu}(\omega(M_f^2,M_{\bar g}^2,\alpha,Q^2))\\
\big[ {\cal C}_0^{\rm iu}(\omega_{fg}(\alpha,P^2))
%
%&& \quad + \, {\cal C}^{\rm iu}_1(\omega(M_f^2,M_{\bar g}^2,\alpha,Q^2))\big],
+ \, {\cal C}^{\rm iu}_1(\omega_{fg}(\alpha,P^2))\big], \label{avwtiP}
\end{equation}
where ($\check \alpha = 1-\alpha$)
\begin{align}
%\omega(M_f^2,M_{\bar g}^2,\alpha,Q^2) &=& M_f^2 \hat \alpha + \alpha M_{\bar g}^2 + \alpha \hat\alpha Q^2\,,
\omega_{fg}(\alpha,P^2) &= M_f^2\check\alpha+\alpha M_g^2+ \alpha \check\alpha P^2\,.
\label{eq:omega}
\end{align}
This is crucial, \emph{e.g}., in proving the axialvector Ward-Green-Takahashi identity.

\subsection{Kaon Bethe-Salpeter amplitude}
The kaon emerges as a quark +anti\-quark bound-state, whose structure is described by a Bethe-Salpeter amplitude.  In the SCI, that amplitude takes the following form:
\begin{align}
\Gamma_{K}(P) = \gamma_5 \left[ i E_{K}(P) + \frac{1}{2M_{fg}}\gamma\cdot P F_{K}(P)\right]\,,
%\Gamma_{0^-}(Q) = \gamma_5 \left[ i E_{0^-} + \frac{1}{M}\gamma\cdot P F_{0^-}\right]\,.
\label{PSBSAA}
\end{align}
where $M_{fg}=M_fM_g/(M_f+M_g)$, $f=u$, $g=\bar s$ for the $K^+$;
$P$ is the kaon total momentum, $P^2 = -m_{K}^2$, $m_K$ is the kaon mass.
As stressed elsewhere \cite{GutierrezGuerrero:2010md, Chen:2012txa}, the axialvector Ward-Green-Takahashi identity is violated of one omits the $\gamma\cdot P F_{K}(P)$ term.
(Pion results are readily obtained by taking $s\to d$.)

The kaon bound-state amplitude and $m_{K}^2$ are obtained by solving the SCI Bethe-Salpeter equation $(t_+ = t+P)$:
\begin{align}
\Gamma_{K}(P)  & =  - \frac{16 \pi}{3} \frac{\alpha_{\rm IR}}{m_G^2}
\int \! \frac{d^4t}{(2\pi)^4} \gamma_\mu S_u(t_+) \Gamma_{K}(P)S_s(t) \gamma_\mu \,.
\label{LBSEI}
\end{align}
Completing appropriate spinor projections, one arrives at the following matrix equation:
\begin{equation}
\label{bsefinalE}
\left[
\begin{array}{c}
E_{K}(P)\\
F_{K}(P)
\end{array}
\right]
= \frac{4 \alpha_{\rm IR}}{3\pi m_G^2}
\left[
\begin{array}{cc}
{\cal K}_{EE}^{K} & {\cal K}_{EF}^{K} \\
{\cal K}_{FE}^{K} & {\cal K}_{FF}^{K}
\end{array}\right]
\left[\begin{array}{c}
E_{K}(P)\\
F_{K}(P)
\end{array}
\right],
\end{equation}
with
{\allowdisplaybreaks
\begin{subequations}
\label{fgKernel}
\begin{align}
\nonumber
{\cal K}_{EE}^{K} &=
\int_0^1d\alpha \bigg\{
{\cal C}_0^{\rm iu}(\omega_{fg}( \alpha, P^2))  \\
& \quad +\left[M_f M_g - \alpha\check\alpha P^2 -\omega_{fg}(\alpha,P^2)\right]\nonumber\\
& \qquad \times\overline{\cal C}^{\rm iu}_1(\omega_{fg}(\alpha, P^2))\bigg\},\\
{\cal K}_{EF}^{K} &= \frac{P^2}{2M_{fg}} \int_0^1d\alpha\, \left[\check\alpha M_f+\alpha M_g\right]\overline{\cal C}^{\rm iu}_1(\omega_{fg}(\alpha, P^2)),\\
{\cal K}_{FE}^{K} &= \frac{2M_{fg}^2}{P^2}{\cal K}_{EF}^{K},\\
{\cal K}_{FF}^{K} &= - \frac{1}{2}\int_0^1d\alpha\, \left[M_f M_g+\hat\alpha M_f^2+\alpha M_g^2\right] \nonumber \\
& \qquad \times \overline{\cal C}^{\rm iu}_1(\omega_{fg}(\alpha, P^2))\,.
\end{align}
\end{subequations}}
\hspace*{-0.2\parindent}The value of $P^2=-m_{K}^2$ for which Eq.\,\eqref{bsefinalE} is satisfied corresponds to the bound-state mass.
The calculated result is listed in Table~\ref{Tab:DressedQuarks}, and the associated solution vector is the kaon's Bethe-Salpeter amplitude.

When calculating observables, the canonically normalised bound-state amplitude must be used, \emph{i.e}., the amplitude obtained after rescaling such that
\begin{equation}
\label{normcan}
1=\left. \frac{d}{d P^2}\Pi_{K}(Z,P)\right|_{Z=P},
\end{equation}
where:
\begin{align}
\Pi_{K}(Z,Q) & = 6 {\rm tr}_{\rm D} \!\! \int\! \frac{d^4t}{(2\pi)^4}   \Gamma_{K}(-Z)
 S(t_+) \, \Gamma_{K}(Z)\, S(t)\,.
 \label{normcan2}
\end{align}
The dimensionless results for $\pi$, $K$ are:
\begin{subequations}\label{kaonBSA}
\begin{align}
E_K & = 3.70\,, \quad F_K = 0.55\,,\\
E_\pi & = 3.59\,, \quad F_\pi = 0.47\,.
\end{align}
\end{subequations}

Using the canonically normalised Bethe-Salpe\-ter amplitude, the kaon leptonic decay constant is
\begin{align}
f_{K} &= \frac{N_c}{4\pi^2}\frac{1}{M_{fg}}\,
\big[ E_{K} {\cal K}_{FE}^{K} + F_{K}{\cal K}_{FF}^{K} \big]_{Q^2=-m_{K}^2}\,. \label{ffg}
\end{align}
%With our normalisation, the empirical value of the kaon's leptonic decay constant is $f_K =0.11\,$GeV \cite{Xu:2021iwv}.

%%\bibliographystyle{elsarticle-num-names}
%%\bibliography{../../../../CollectedBiB}

\end{document}